\documentclass[12pt]{article}
\usepackage{amssymb,amsmath,epsfig}

\begin{document}

\title{\textbf{Anisotropic Compact Stars in $f(T)$ Gravity}}
\medskip
\author{G. Abbas$^a$ \thanks{
ghulamabbas@ciitsahiwal.edu.pk}, Afshan Kanwal $^a$
\thanks{afshan@ciitsahiwal.edu.pk} and M. Zubair $^b$\thanks
{mzubairkk@gmail.com; drmzubair@ciitlahore.edu.pk}\\
$^{a}$Department of Mathematics, COMSATS Institute of
\\Information
Technology Sahiwal-57000, Pakistan \\
$^{b}$Department of Mathematics, COMSATS Institute of
\\Information
Technology Lahore, Pakistan}
\date{}
\maketitle
\begin{abstract}
This paper deals with the theoretical modeling of anisotropic
compact stars in the framework of $f(T)$ theory of gravity, where
$T$ is torsion scalar. To this end, we have used the exact solutions
of Krori and Barua metric to a static spherically symmetric metric.
The unknown constants involved in the Krori and Barua metric have
been specified by using the masses and radii of compact stars
4$U$1820-30, Her X-1, SAX J 1808-3658. The physical properties of
these stars have been analyzed in the framework of $f(T)$ theory. In
this setting, we have checked the anisotropic behavior, regularity
conditions, stability and surface redshift of the compact stars.
\end{abstract}


\section{Introduction}

On the basis of Einstein's proposal for the investigation of a new
version to General Relativity (GR) (Unzicker \& Case), an
alternative theory of gravitation, termed as Teleparallel Theory
(TT) has been adopted many years later from the original formulation
of GR. However, the analogy between GR and TT has been tackled once
again on the basis of postulates defined by Moller (Moller 1961;
Pellegrini \& Plebanski 1963; Moller 1978; Hayashi \& Takano 1967;
Hayashi 1973, 1977). According to GR the gravitational effects due
to a gravitating source can be described by the curvature of that
source. In general it is true that a spacetime may posses curvature
and torsion (as in case of Cartan space), one can distinguish all
the terms resulting from torsion of spacetime as Riemann tensor,
connection, etc...Therefore, one can remark that the theory that
accounts gravity as an action of curvature of spactime (resulting
from the Riemann tensor without either torsion or antisymmetric
connection) can be considered as a theory that consists of only
torsion with null contribution from Riemann tensor without torsion
bec (Capozziello \& Faraoni 2011).

On the basis of current expanding paradigm and existence of exotic
energy component named as dark energy, various modifications of GR
have been proposed. The unified theories of gravitation in
low-energy scales involves the terms $R^{2},R^{\mu\nu} R_{\mu\nu}$
and $R^{\mu\nu\alpha\beta}R_{\mu\nu\alpha\beta}$, in their effective
actions. In such candidates $f(R)$ has gained great interest and it
agrees with the cosmological and astrophysical observational data
(Nojiri \& Odintsov 2011). In $f(R)$ gravity, the dynamical
equations are of the fourth order differential equations, so it
becomes more difficult to solve these equations as compared to GR
(Nojiri \& Odintsov 2007). As the GR has the similarity with
Teleparallel theory, a theory namely $f(T)$ (where $T$ is a torsion
scalar) has been used. Such theory would be the alternative form of
the generalization of GR, namely the $f(R)$ theory of gravity. This
newly proposed $f(T)$ theory is the modification of TT, which is
free of curvature and Riemann tensors resulting from the terms
without torsion.

In the relativistic cosmology, $f(T)$ theory has been employed to
discuss the inflation of the universe (Ferraro \& Fiorini 2007).
Since then there has been growing interest to study the various
aspects of cosmology in $f(T)$ theory of gravity. Many authors
(Hayashi 1977; Capozziello \& Faraoni 2011; Nojiri \& Odintsov 2007,
2011; Ferraro \& Fiorini 2007; Bayin 1982; Linder 2001; Bamba et al.
2011; Zubair \& Waheed 2014), investigated that this theory can be
utilized as a handy candidate for the accelerated expansion of the
universe without the inclusion of dark energy. In the astronomy and
astrophysics, the $f(T)$ theory was initially used to drive the BTZ
black hole solutions (Dent et al. 2011). Bamba et al. (2011) and
Miao et al. (2011) shown that that first of black hole
thermodynamics does not hold in $f(T)$ ttheory. Recently, some
static spherically symmetric solutions with Maxwell term have been
found in $f(T)$ theory (Wang 2011). Boehmer et al. (2011) and Daouda
(2011) investigated the existence of relativistic stars in $f(T)$
theory developing the various static perfect fluid solutions.

In this paper, we formulate the models of anisotropic compact
objects in $f(T)$ theory, without using the equation of state. We
have discussed the various properties of the compact objects. This
paper is organized as follows: The brief review of Weitzenbock's
geometry and equation of motion of $f(T)$ gravity will be presented
in section \textbf{2}. Section \textbf{3}, deals with the geometry
of the source and physical significance of matter component. The
physical analysis of the proposed model is presented in section
\textbf{4}. Finally, we discuss the summary of results in the last
section.

\section{$f(T)$ Theory of Gravity}

Recently, it has been proved that TT is equivalent theory to the GR
(Bayin 1982; Linder 2001; Rehaman 2010). Here, we briefly introduce
the basic concept of TT, for this purpose, it is assumed that Latin
and Greek indices are related to the tetrad fields and spacetime
coordinates, respectively. In this case metric of spacetime
spacetime is defined as follows:
\begin{equation}
ds^{2}=g_{\mu\nu}dx^{\mu}dx^{\nu},
\end{equation}
The above metric can be transformed to Minkowskian description by
the tetrad matrix,defined by
\begin{equation}
dS^{2}=g_{\mu\nu}dx^{\mu}dx^{\nu}=\eta_{ij}\theta^{i}\theta^{j},
\end{equation}
\begin{equation}
dx^{\mu}=e_{i}^{\mu}\theta^{i}, \theta^{i}=e^{i}_{{\mu}}dx^{\mu},
\end{equation}
where $\eta_{ij}=diag[1,-1,-1,-1]$ and
$e_{i}^{\mu}e_{i}^{\nu}=\delta_{\nu}^{\mu}$ or
$e_{i}^{\mu}e_{j}^{\nu}=\delta_{i}^{j}$. \\
The root of the metric determinant is given by $\sqrt{-g
}=det[e_{\mu}^{i}]=e. $ The Weitzenbock's connection components for
vanishing  Riemann tensor part and non-vanishing torsion term are
defined as
\begin{equation}
\Gamma^{\alpha}_{\mu\nu}=e_{i}^{\alpha}\partial_{\nu}e_{\mu}^{i}=-e_{i}^{\mu}\partial_{\nu}e_{i}^{\alpha}
\end{equation}
The torsion and the contorsion are defined by
\begin{equation}
T^{\alpha}_{\mu\nu}=\Gamma^{\alpha}_{\nu\mu}-\Gamma^{\alpha}_{\mu\nu}=e_{i}^{\alpha}(\partial_{\mu}e_{\nu}^{i}-\partial_{\nu}e_{\mu}^{i})
\end{equation}
\begin{equation}
K^{\mu\nu}_{\alpha}=-\frac{1}{2}(T^{\mu\nu}_{\alpha}-T^{\nu\mu}_{\alpha}-T^{\mu\nu}_{\alpha})
\end{equation}
and the components of the tensor $S_{\alpha}^{\mu\nu}$ as
\begin{equation}
S_{\alpha}^{\mu\nu}=\frac{1}{2}(K^{\mu\nu}_{\alpha}+\delta^{\mu}_{\alpha}T^{\beta\nu}_{\beta}-\delta^{\nu}_{\alpha}T^{\beta\mu}_{\beta}).
\end{equation}
Here, torsion scalar is
\begin{equation}
T=T^{\alpha}_{\mu\nu}S_{\alpha}^{\mu\nu}
\end{equation}
Analogous to $f(R)$ gravity, the action for $f(T)$ gravity is
\begin{equation}
S[e_{\mu}^{i}, \Phi_A]=\int d^4 x
e[\frac{1}{16\pi}f(T)+\mathcal{L}_Matter(\Phi_A)],
\end{equation}
in the above action $G=c=1$ have been used and the
$\mathcal{L}_{Matter}(\Phi_A)$ is matter field. The variation of the
above action provide the following field equations in $f(T)$ gravity
(Dent et al. 2011)
\begin{equation}
S_{\mu}^{\nu\rho}\partial_{\rho}T
f_{TT}+[e^{-1}e^{i}_{\mu}\partial_{\rho}(ee_i^{\alpha}S_{\alpha}^{\nu\rho})+T^{\alpha}_{\lambda\mu}
S_{\alpha}^{\nu\lambda}]f_T+\frac{1}{4}\delta^{\nu}_{\mu}f=4\pi
\mathcal{T}_{\mu}^{\nu},
\end{equation}
where $\mathcal{T}_{\mu}^{\nu}$ is matter. In the present case, we
take the matter as anisotropic fluid for which energy-momentum
tensor is
\begin{equation}
\mathcal{T}_{\mu}^{\nu}=(\rho+p_{t})u_{\mu}u^{\nu}-p_{t}\delta_{\mu}^{\nu}+(p_{r}-p_{t})v_{\mu}v^{\nu},
\end{equation}
where $u^{\mu}$ and  $v^{\mu}$ are the four-velocity and radial-four
vectors, respectively. Further, $p_r$ and $p_t$ are pressures along
radial and tansverse directions.

\section{Model of Anisotropic Compact Stars in Generalized Telleparallel Gravity}

We assume the geometry of star in the form of static spherically
symmetric spacetime which is given by
\begin{equation}\label{3.1}
ds^{2}=e^{a(r)}dt^{2}-e^{b(r)}dr^{2}-r^{2}(d\theta^{2}+sin^{2}(\theta)d\phi^{2}).
\end{equation}
 We introduce the tetrad matrix for (\ref{3.1}) as follows:
\begin{equation}\label{3.2}
[e_{\mu}^{i}]=diag[e^\frac{a(r)}{2},e^\frac{b(r)}{2},r,r
sin(\theta)].
\end{equation}
Using Eq.(13), one can obtain
$e=det[e_{\mu}^{i}]=e^\frac{(a+b)}{2}r^{2}sin (\theta)$, and with
(4)-(8), torsion scalar and its derivative are determined in terms
of $r$ as
\begin{eqnarray}\label{3.3}
T(r)=\frac {2 e^{-b}}{r}(a^{'}+\frac {1}{r}), \\\label{3.4}
T^{'}(r)=\frac {2 e^{-b}}{r}(a^{''}+\frac
{1}{r^{2}}-T(b^{'}+\frac{1}{r})),
\end{eqnarray}
where prime represents derivative with respect to $r$. The set of
equations for an anisotropic fluid as
\begin{eqnarray}\label{3.5}
4\pi\rho=\frac {f}{4}-\frac{f_{T}}{2}\left(T-\frac{1}{r^{2}}-\frac {
e^{-b}}{r}(a^{'}+b^{'})\right) ,\\\label{3.6}
 4\pi
p_{r}=\frac{f_{T}}{2}\left(T-\frac{1}{r^{2}}\right)-\frac
{f}{4},\\\label{3.7}
 4\pi
p_{t}=\left[\frac{T}{2}+e^{-b}\left(\frac{a^{''}}{2}+(\frac
{a^{'}}{4}+\frac {1}{2r})(a^{'}-b^{'})\right)\right]\frac
{f_{T}}{2}-\frac {f}{4},\\\label{3.8}
 \frac
{cot\theta}{2r^{2}}T^{'}f_{TT}=0,
\end{eqnarray}
 In the Eqs.(\ref{3.5})-(\ref{3.7}), the Eq.
(\ref{3.8}), has been used. Further, Eq. (\ref{3.8}) leads to the
the following linear form of $f(T)$ :
\begin{equation}\label{3.9}
f(T)=\beta T+{\beta}_{1},
\end{equation}
where $\beta$ and ${\beta}_1$ are integration constants. We
parameterize the metric as the following:
\begin{eqnarray}\label{3.9a}
b(r)=Ar^2,\quad a(r)=Br^2+C
\end{eqnarray}
where arbitrary constants $A$, $B$ and $C$ can be evaluated by using
some physical matching conditions. Now using
Eq.(\ref{3.9}),(\ref{3.9a}), we get following form of matter
components:
\begin{eqnarray}\label{3.10}
&&\rho=-\frac{e^{-Ar^{2}}\beta}{8\pi
r^{2}}+\frac{\beta_{1}}{16\pi}+\frac{\beta}{8\pi
r^{2}}+\frac{e^{-Ar^{2}}A\beta}{4\pi},\\\label{3.11}
&&p_{r}=\frac{e^{-Ar^{2}}B\beta}{4\pi}+\frac{e^{-Ar^{2}}\beta}{8\pi
r^{2}}-\frac{\beta}{8\pi
r^{2}}-\frac{{\beta}_{1}}{16\pi},\\\label{3.12}
&&p_{t}=\frac{e^{-Ar^{2}}B\beta}{4\pi}+\frac{e^{-Ar^{2}}\beta
r^{2}B^{2}}{8\pi}-\frac{AB\beta r^{2}e^{-Ar^{2}}}{8\pi}-
\frac{e^{-Ar^{2}}A\beta}{8\pi}-\frac{{\beta}_{1}}{16\pi},
\end{eqnarray}
Also, the equation of state (EOS) parameters can be written as
\begin{eqnarray}\label{3.13}
&&{\omega}_{r}(r)=\frac{4r^{2}B\beta e^{-Ar^{2}}+2\beta
e^{-Ar^{2}}-2\beta-r^{2}{\beta}_{1}}{-2\beta
e^{-Ar^{2}}+r^{2}{\beta}_{1}+2\beta+4r^{2}A\beta},\\\label{3.14}
&&{\omega}_{t}(r)=\frac{r^{2}B\beta+2r^{4}B^{2}\beta
e^{-Ar^{2}}-2\beta ABr^{4}+2r^{2}A\beta
e^{-Ar^{2}}-4r^{2}{\beta}_{1}}{e^{-Ar^{2}}
+2{\beta}_{1}r^{2}+\beta+2r^{2}A\beta e^{-Ar^{2}}},
\end{eqnarray}

\begin{figure}
\center\epsfig{file=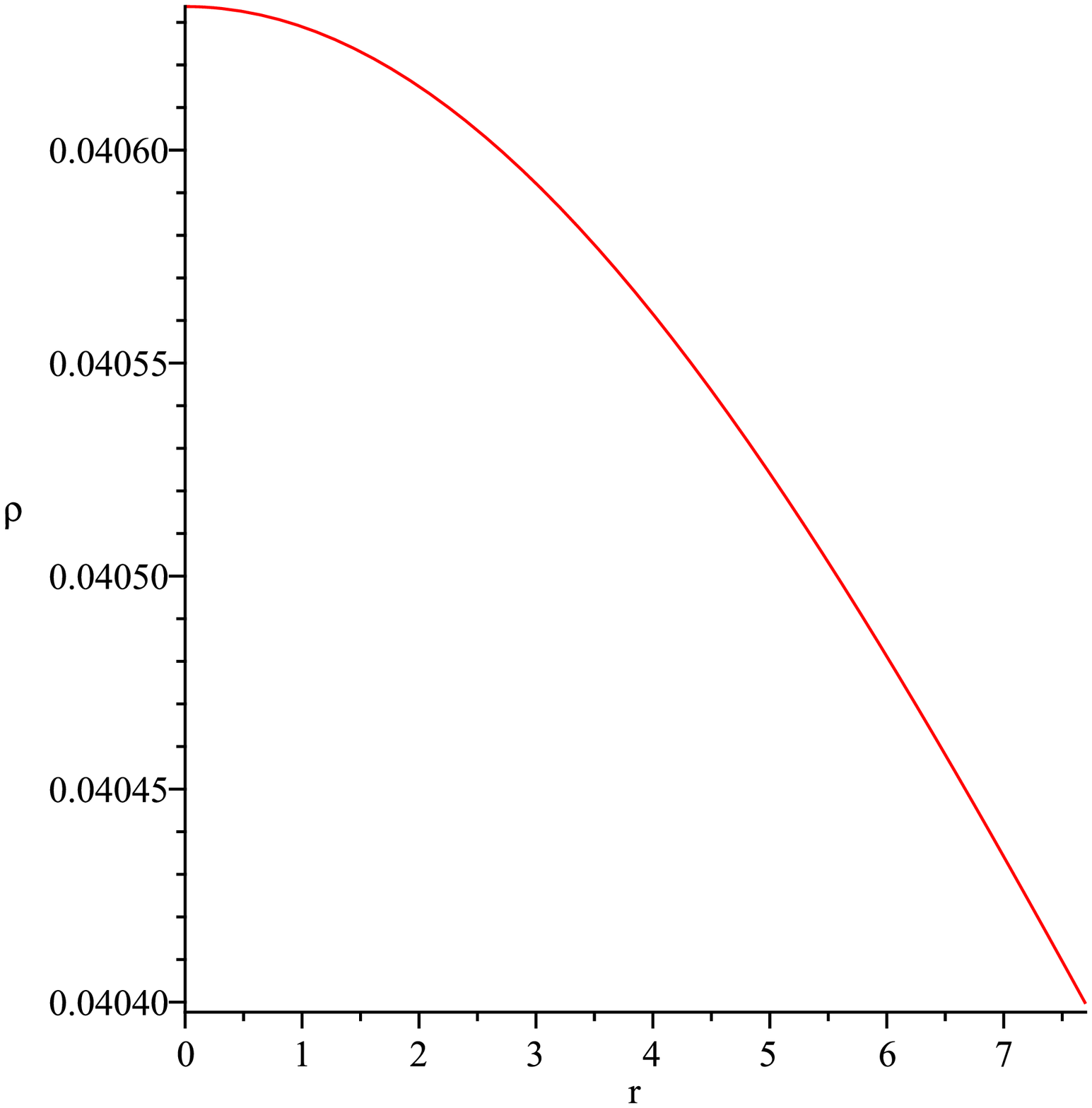, width=0.3\linewidth}
\epsfig{file=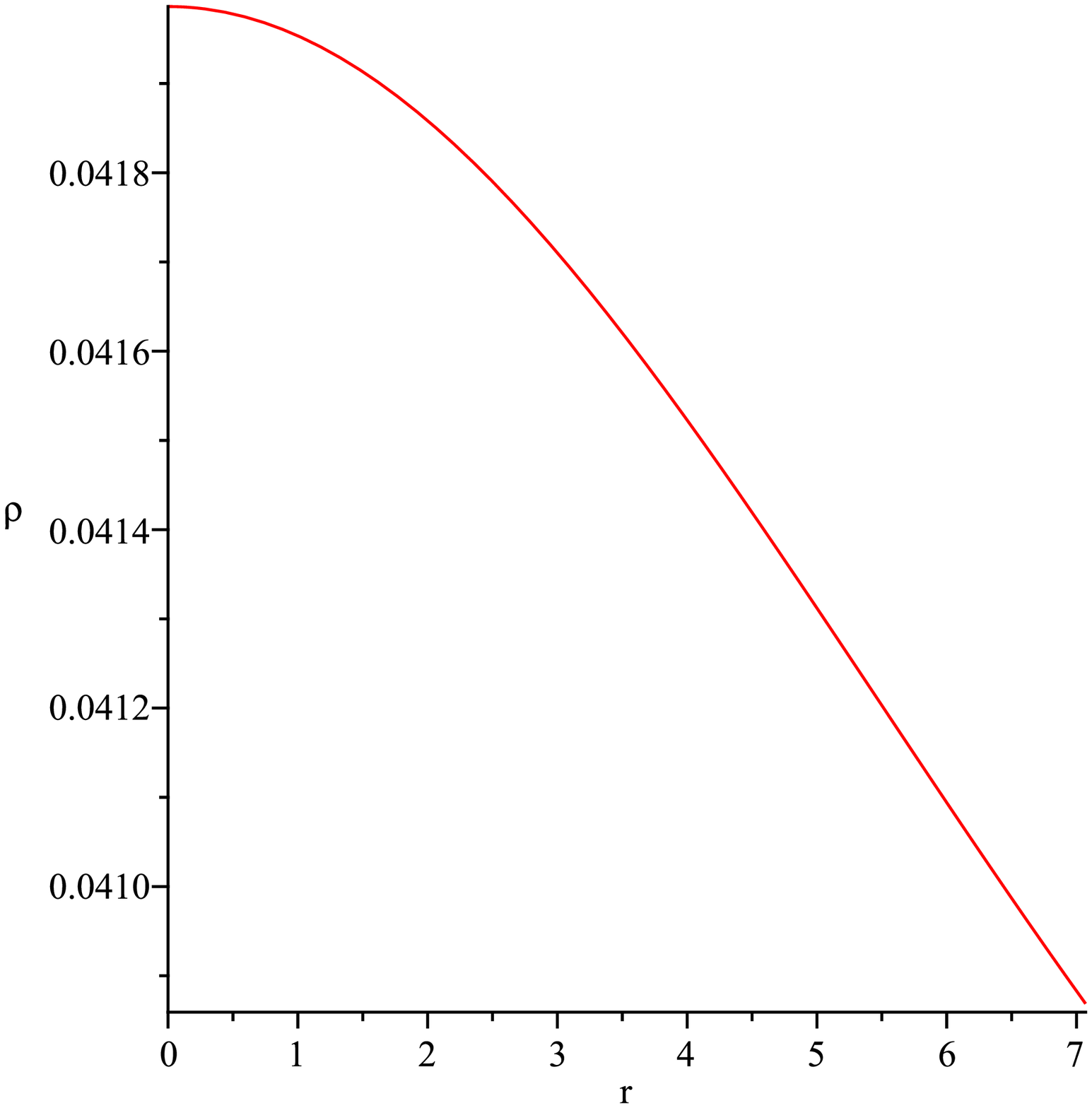, width=0.3\linewidth} \epsfig{file=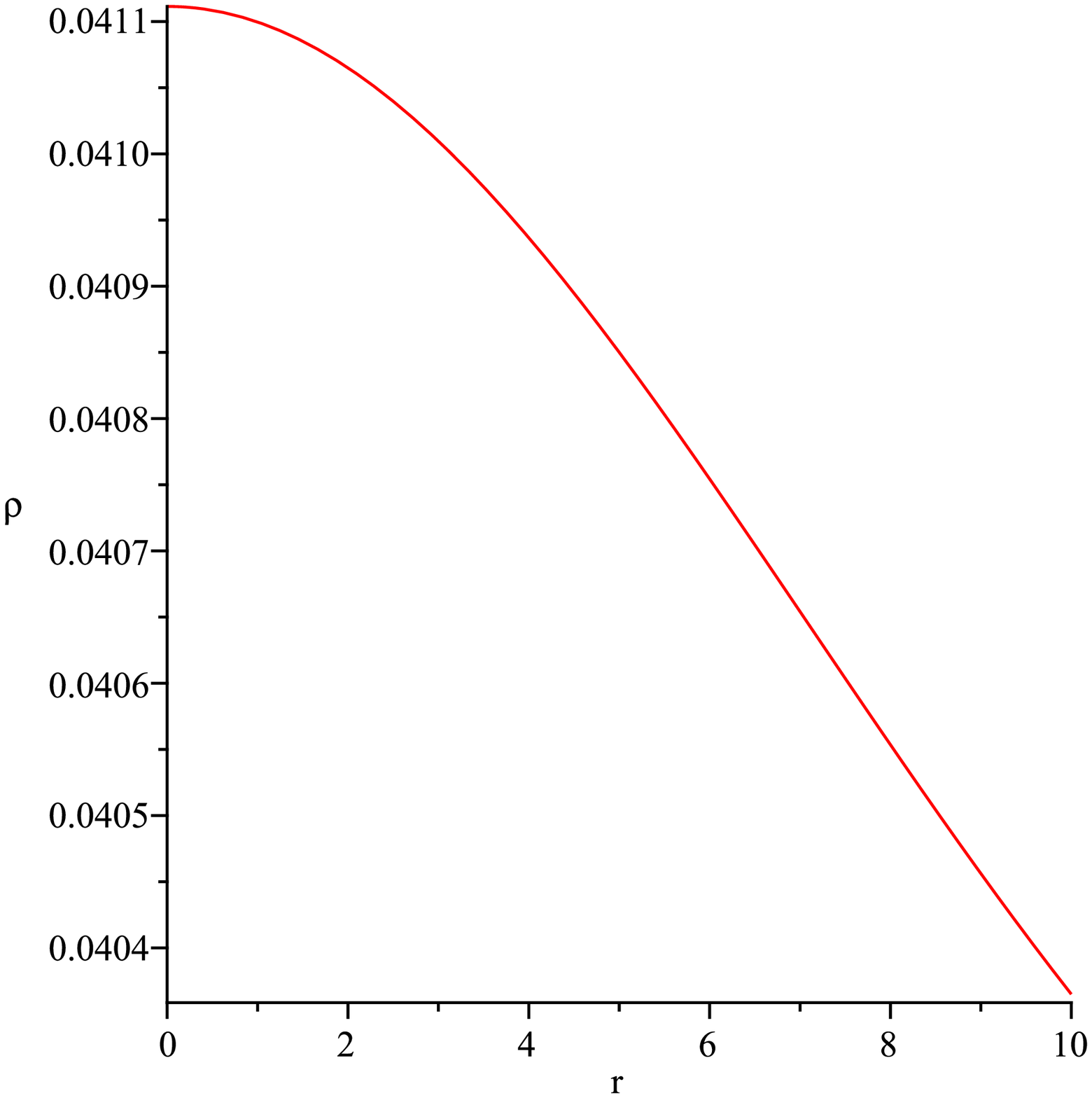,
width=0.3\linewidth}\caption{First, second and third graphs
represent the density variation of Strange star candidate Her X-1,
SAX J 1808.4-3658(SS1) and 4U 1820 - 30, respectively.}
\end{figure}

\begin{figure}
\center\epsfig{file=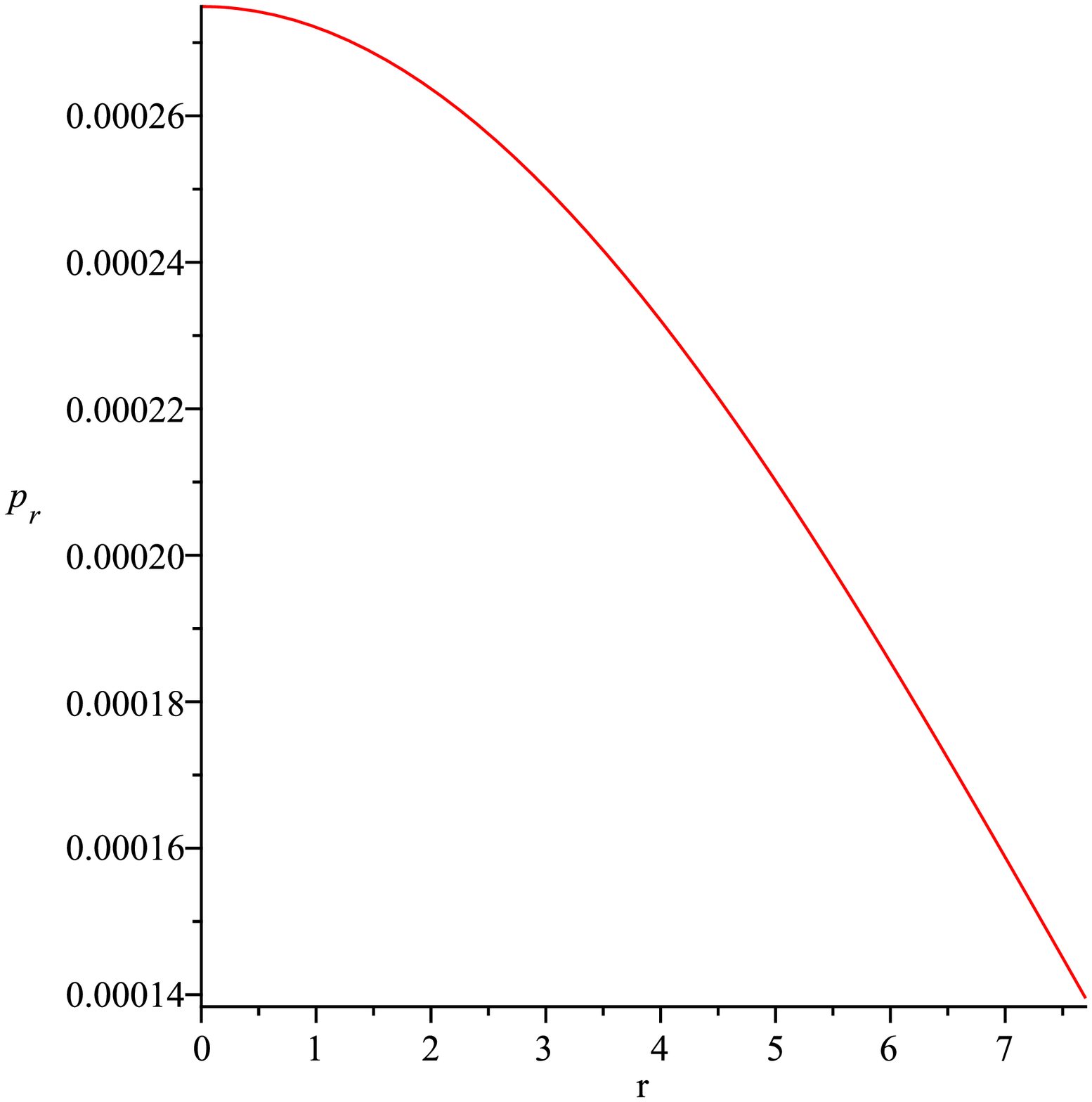, width=0.3\linewidth}
\epsfig{file=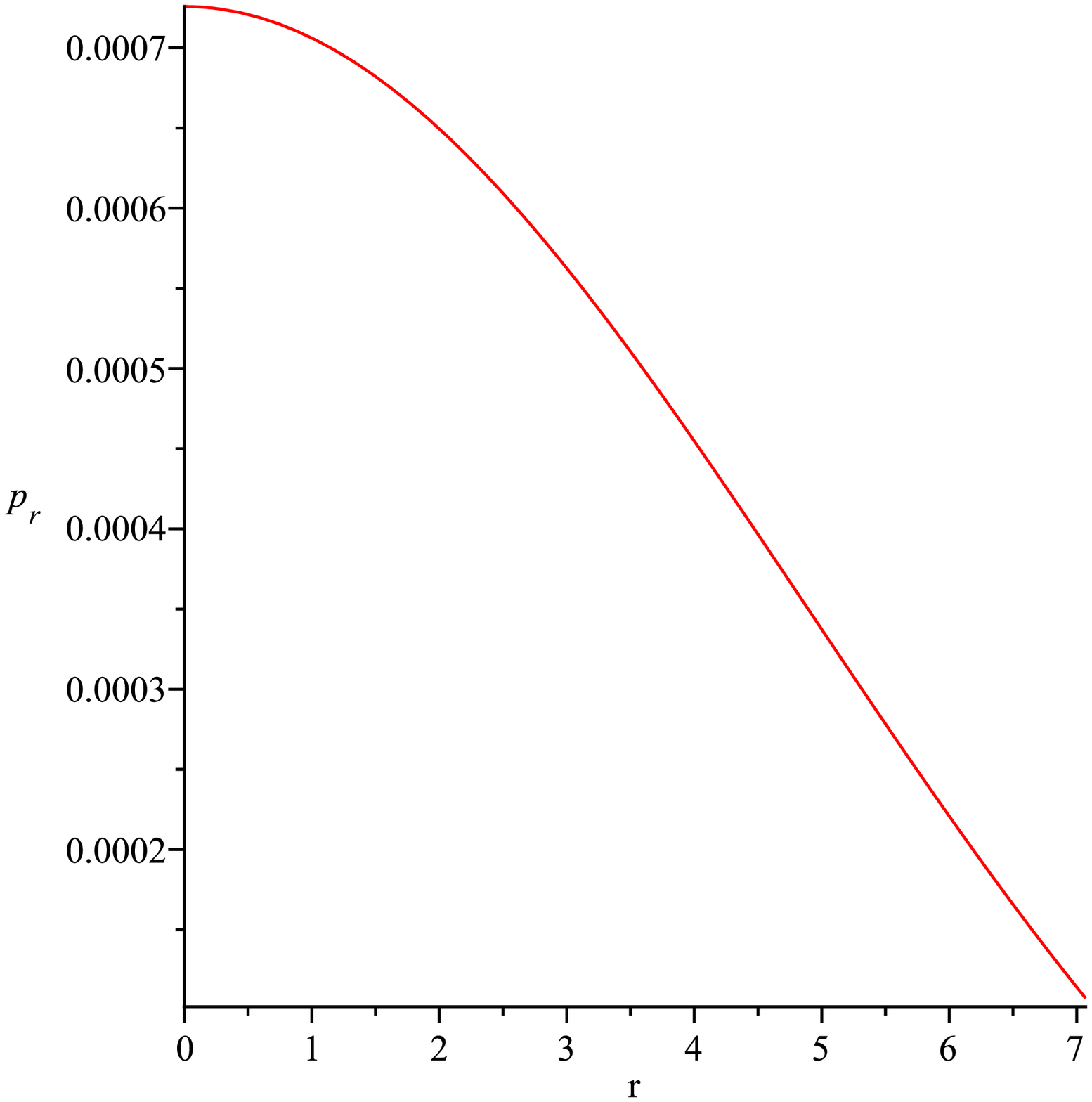, width=0.3\linewidth} \epsfig{file=fig2a.eps,
width=0.3\linewidth}\caption{First, second and third graphs
represent the radial pressure variation of Strange star candidate
Her X-1, SAX J 1808.4-3658(SS1) and 4U 1820 - 30, respectively.}
\end{figure}

\begin{figure}
\center\epsfig{file=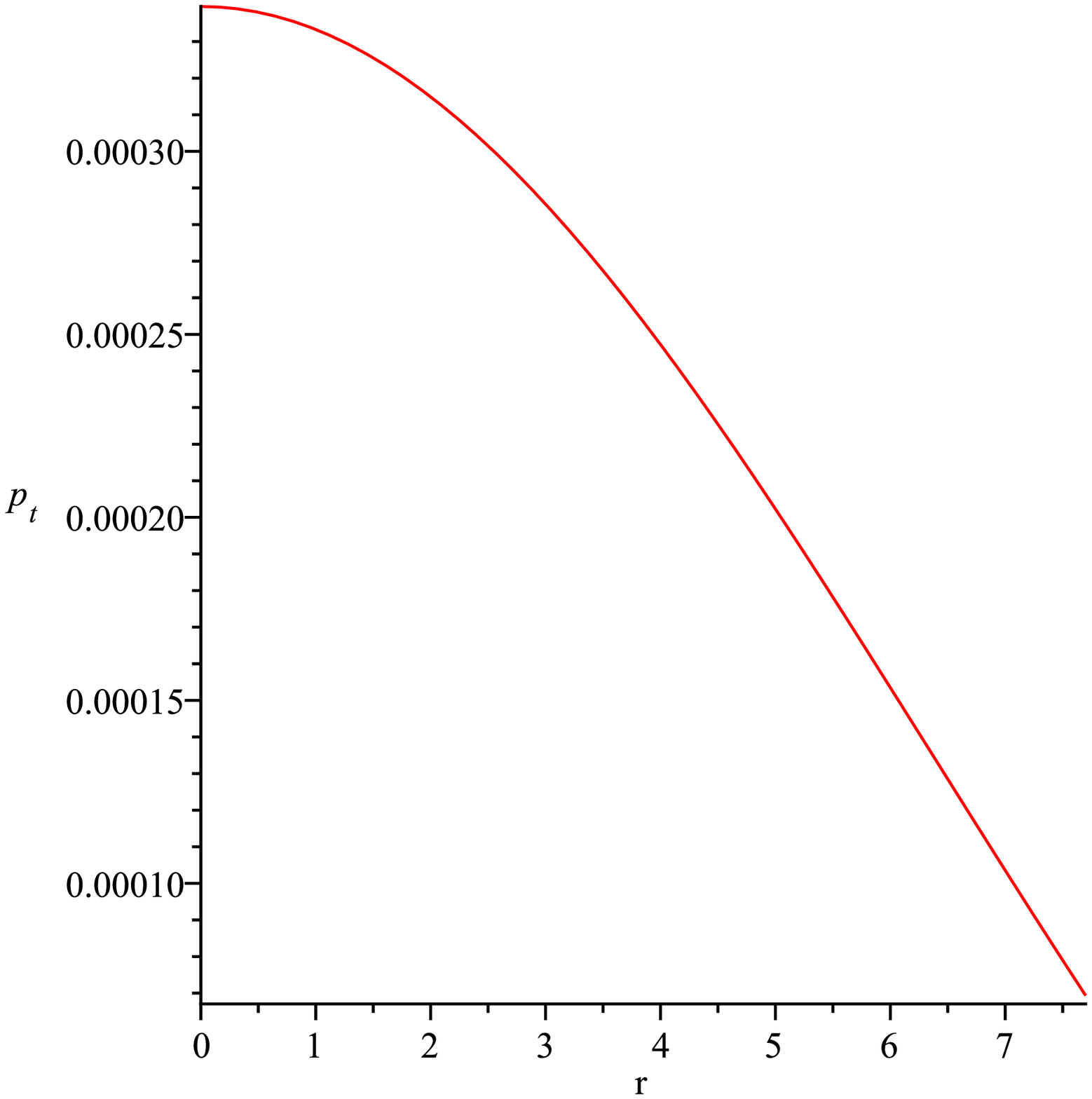, width=0.3\linewidth}
\epsfig{file=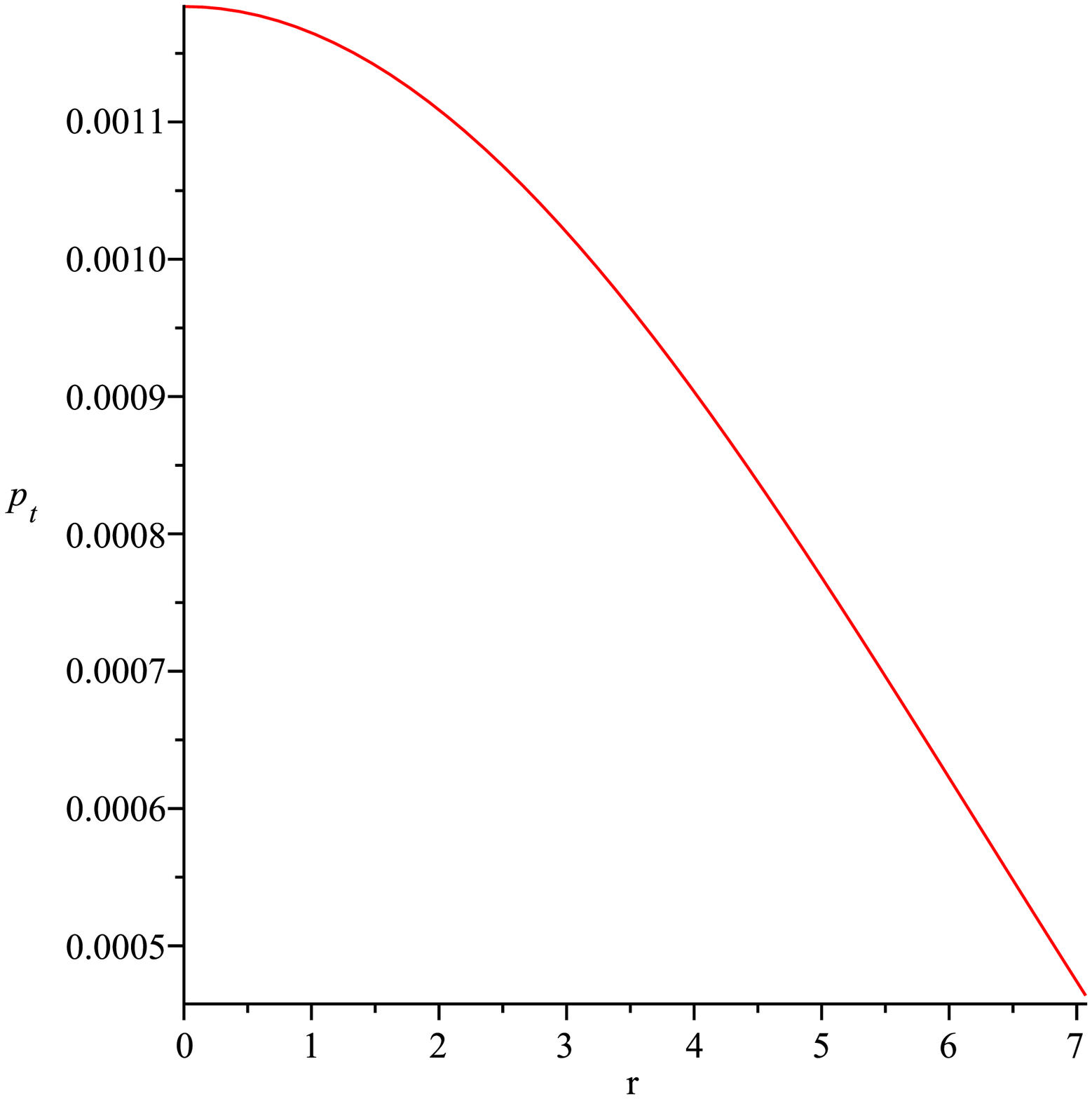, width=0.3\linewidth} \epsfig{file=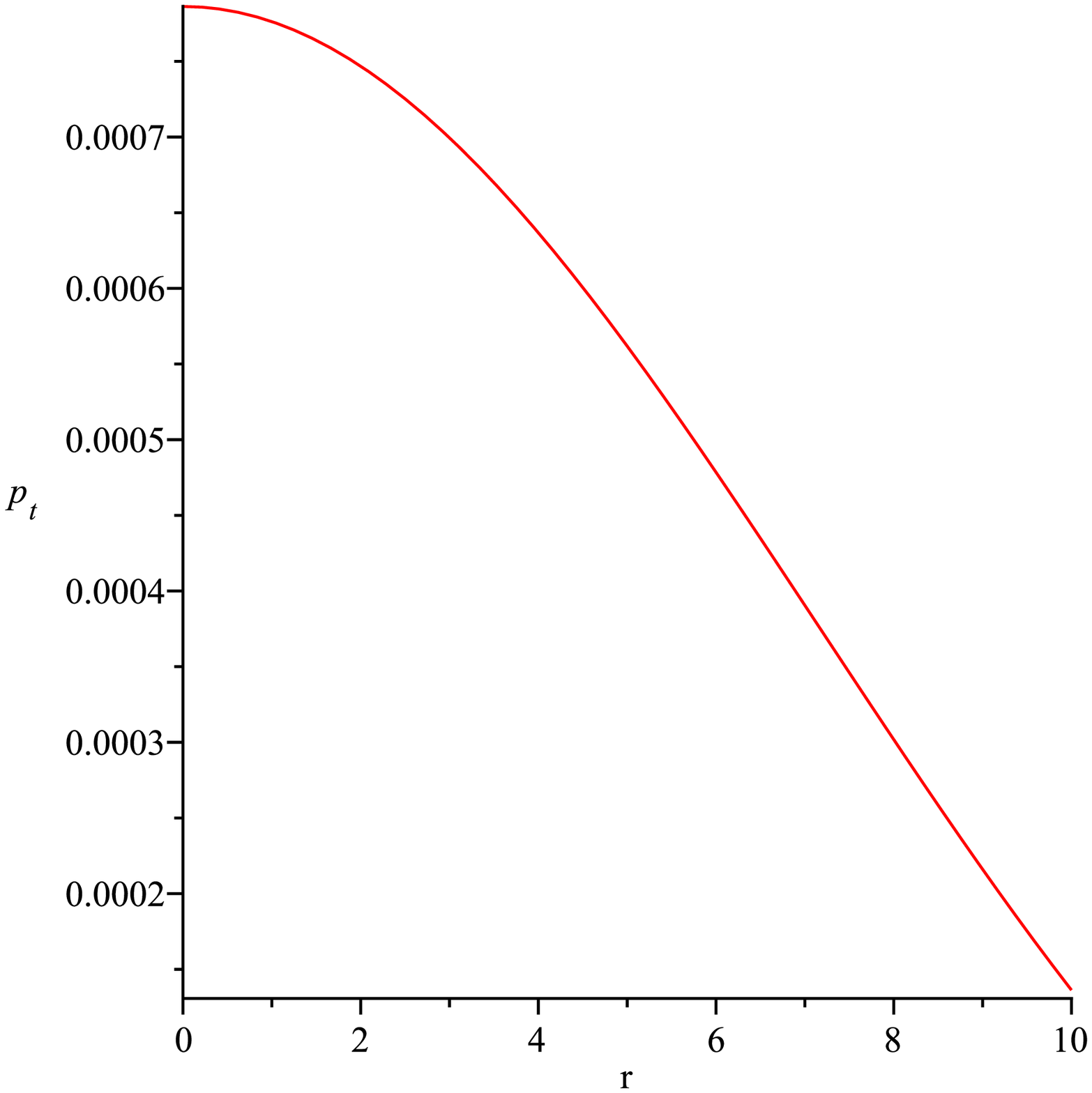,
width=0.3\linewidth}\caption{First, second and third graphs
represent the transverse pressure variation of Strange star
candidate Her X-1, SAX J 1808.4-3658(SS1) and 4U 1820 - 30,
respectively.}
\end{figure}
\begin{figure}
\center\epsfig{file=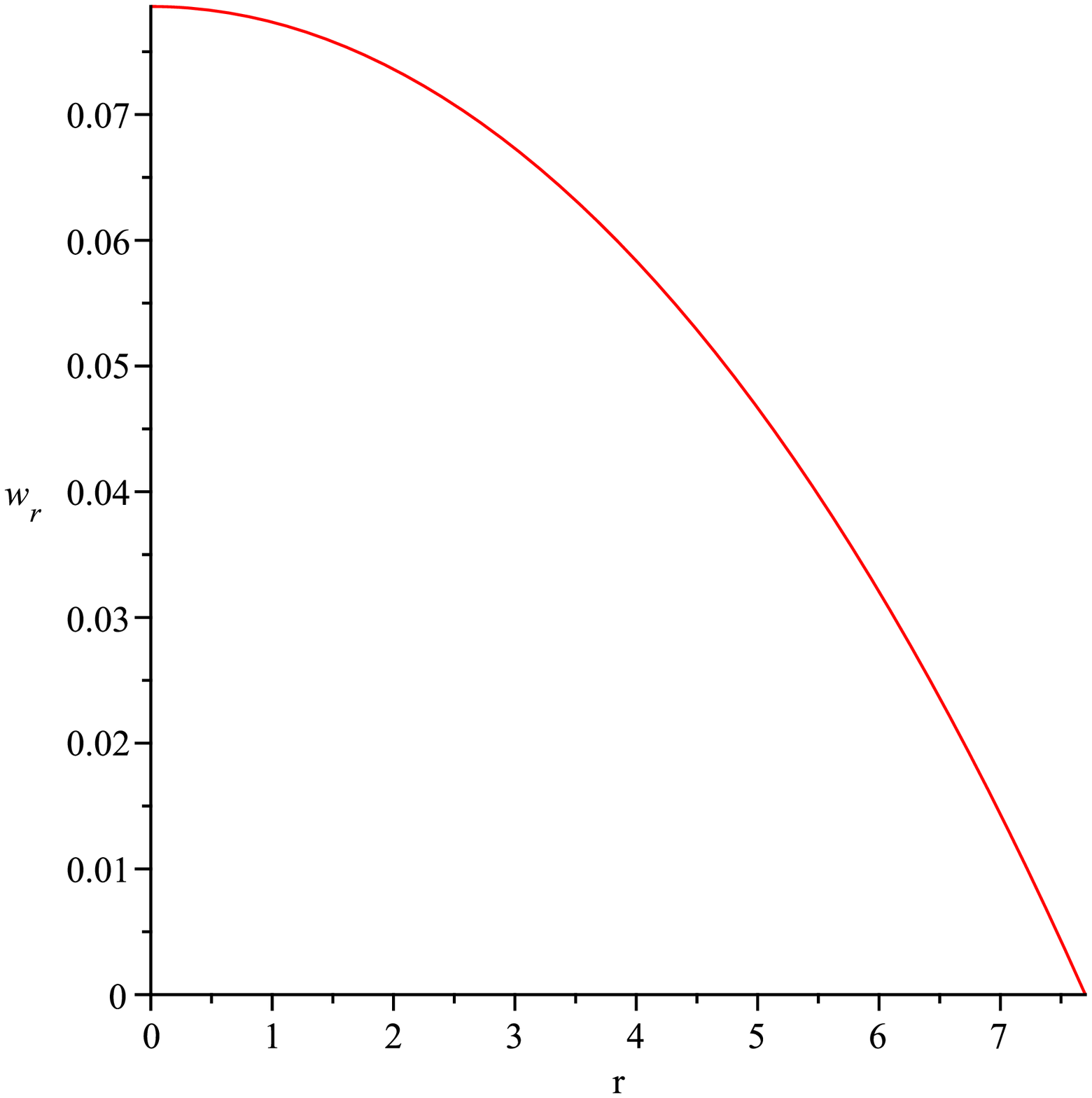, width=0.3\linewidth}
\epsfig{file=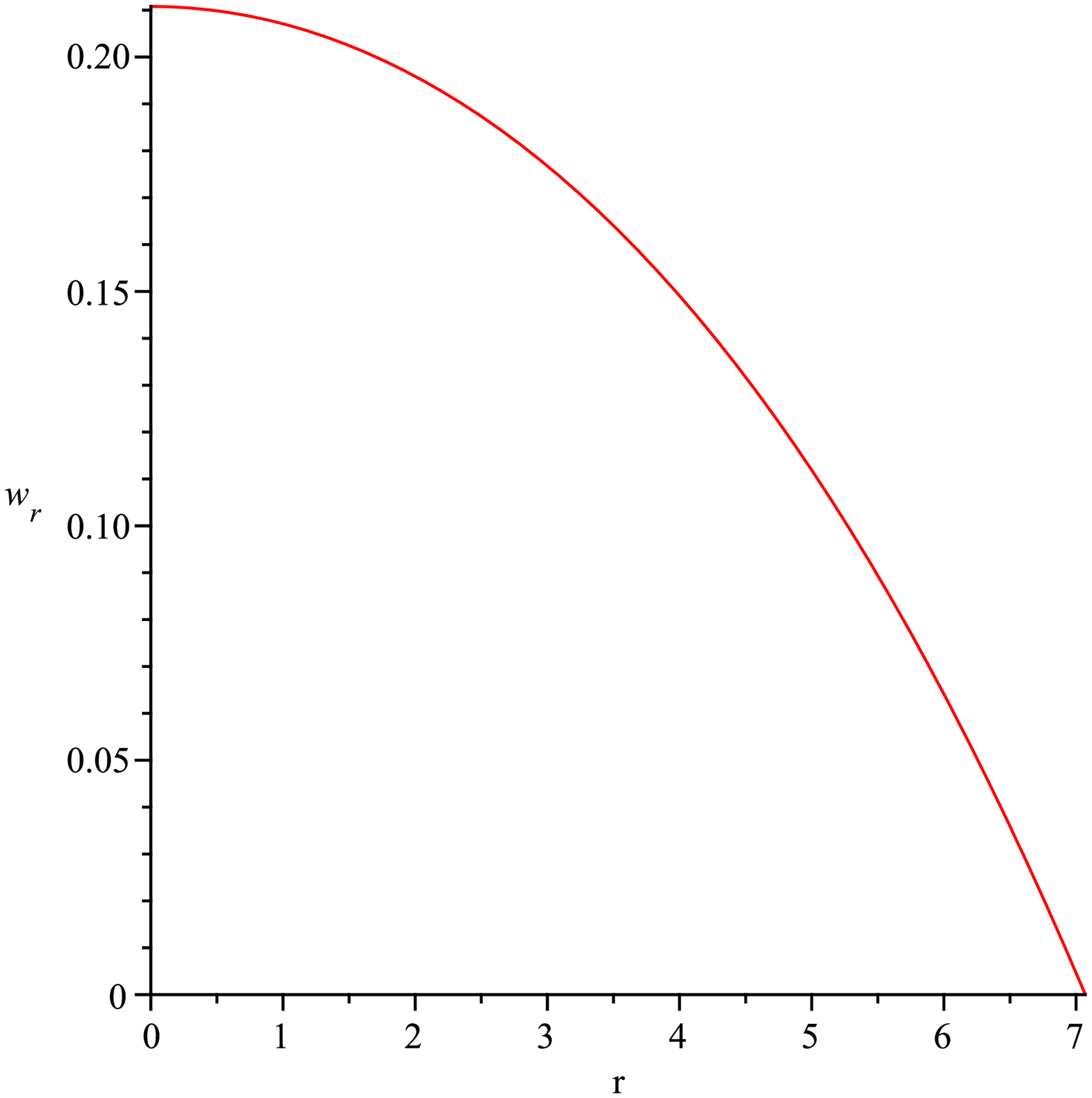, width=0.3\linewidth} \epsfig{file=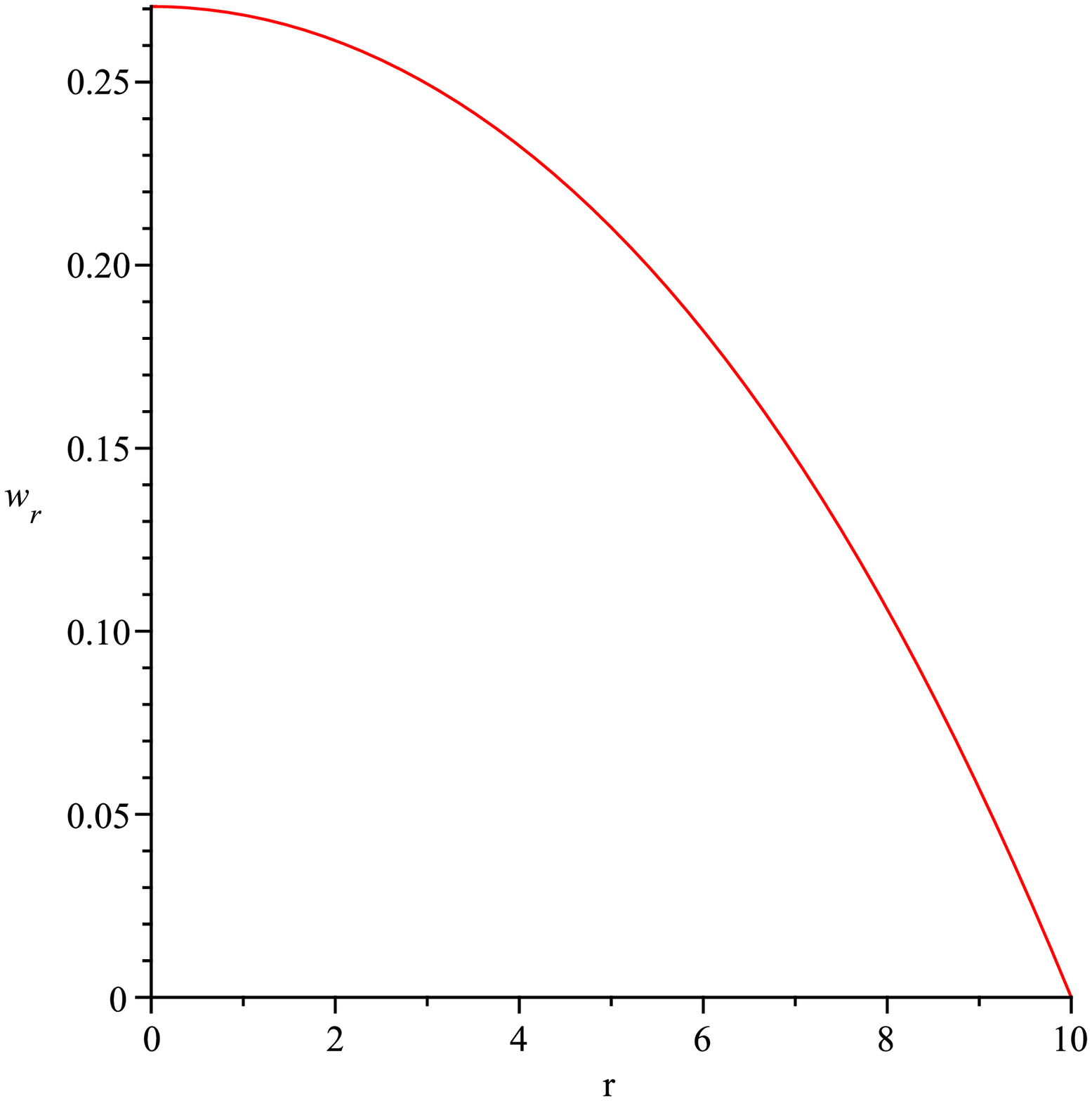,
width=0.3\linewidth}\caption{First, second and third graphs
represent the EOS parameter ${\omega}_{r}$ variation of Strange star
candidate Her X-1, SAX J 1808.4-3658(SS1) and 4U 1820 - 30,
respectively.} \center\epsfig{file=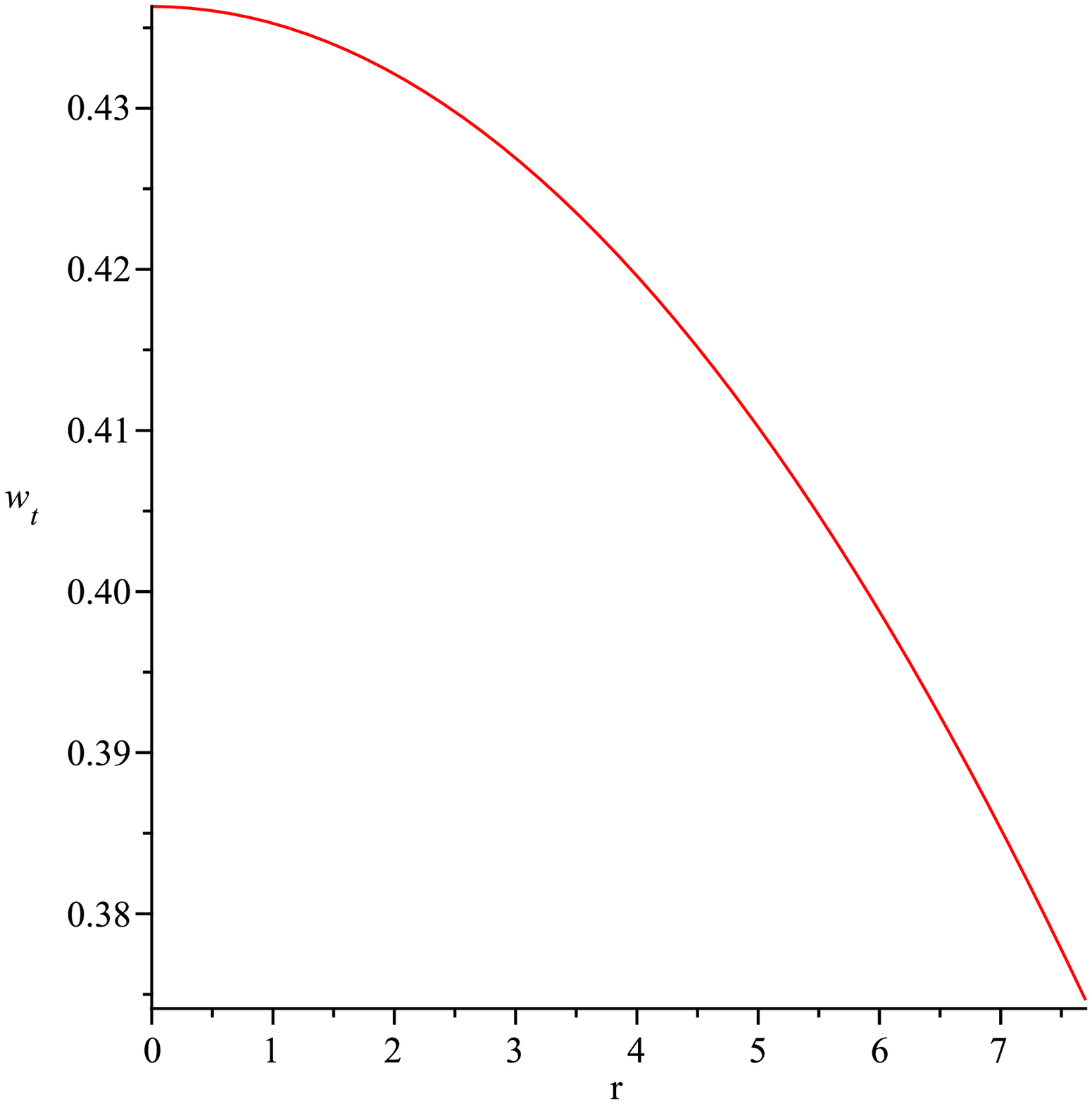, width=0.3\linewidth}
\epsfig{file=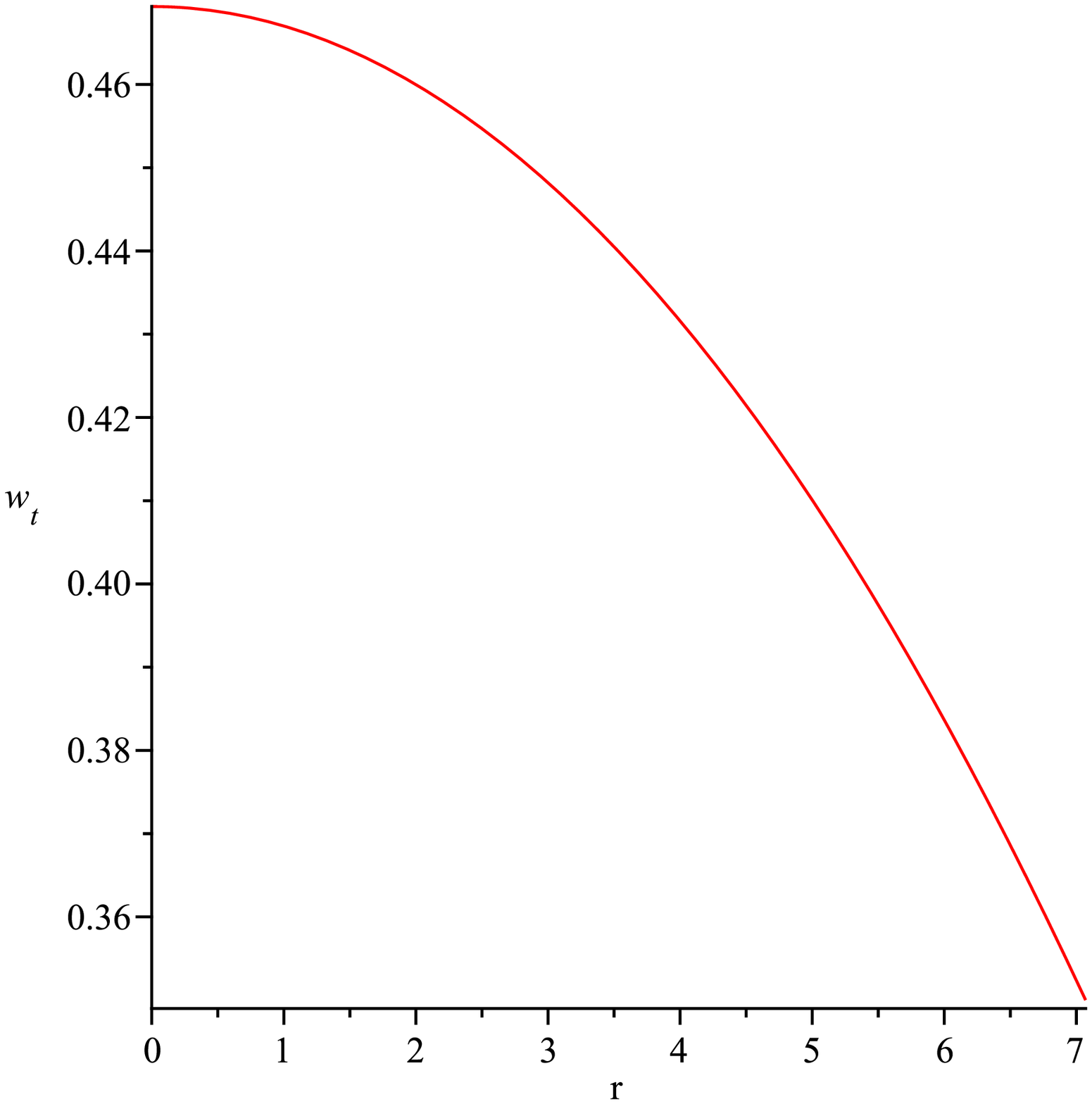, width=0.3\linewidth} \epsfig{file=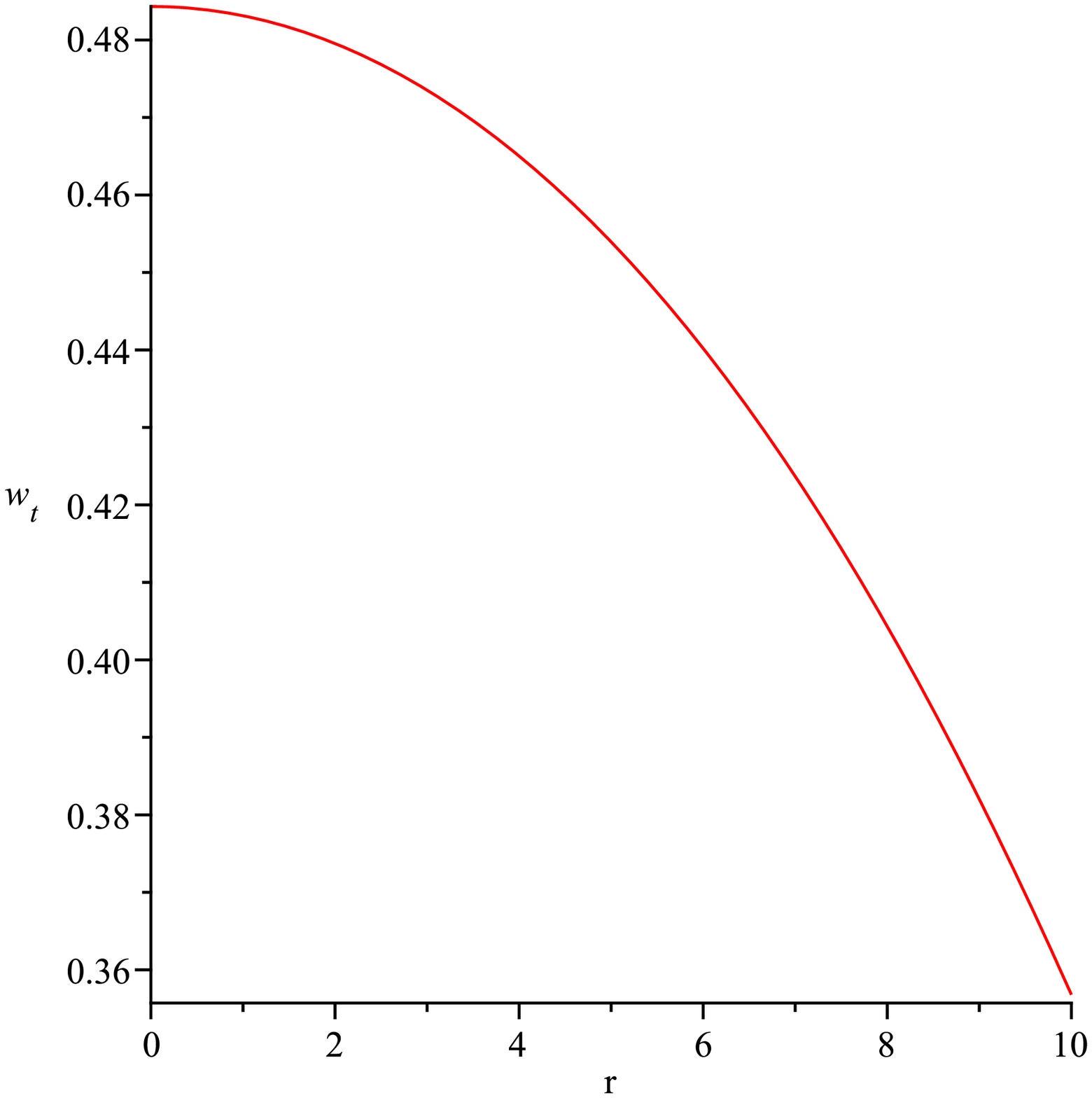,
width=0.3\linewidth}\caption{First, second and third graphs
represent the EOS parameter ${\omega}_{t}$ variation of Strange star
candidate Her X-1, SAX J 1808.4-3658(SS1) and 4U 1820 - 30,
respectively.}
\end{figure}

\section{ Analysis of the Proposed Model}
Here, we discuss the following properties of the proposed model:
\subsection{Anisotropic Behavior}
Using Eqs.(\ref{3.10}) and (\ref{3.11}), we get
\begin{eqnarray}\label{3.10a}
&&\frac{d\rho}{dr}=\frac{\beta Ae^{-Ar^{2}}}{4\pi r}+\frac{\beta
e^{-Ar^{2}}}{4\pi r^{3}}-\frac{\beta}{4\pi r^{3}}-\frac {A^{2}\beta
re^{-Ar^{2}}}{2\pi},\\\label{4.2} &&\frac{dp_{r}}{dr}=-\frac{\beta
ABre^{-Ar^{2}}}{2\pi}-\frac{\beta Ae^{-Ar^{2}}}{4\pi r}-\frac{\beta
e^{-Ar^{2}}}{4\pi r^{3}}+\frac {\beta}{4\pi r^{3}}.
\end{eqnarray}
The above results lead to following equations
\begin{eqnarray}\nonumber
\frac{d^{2}\rho}{dr^{2}}&&=\frac{\beta
A^{2}e^{-Ar^{2}}}{2\pi}-\frac{\beta Ae^{-Ar^{2}}}{4\pi
r^{2}}-\frac{\beta Ae^{-Ar^{2}}}{2\pi r^{2}}-\frac {3\beta
e^{-Ar^{2}}}{4\pi r^{4}}+\frac{3\beta}{4 \pi r^{4}}-\frac {\beta
A^{2}e^{-Ar^{2}}}{2\pi}\\\nonumber&&+\frac{\beta
A^{3}r^{2}e^{-Ar^{2}}}{\pi},\\\nonumber
\frac{d^{2}p_{r}}{dr^{2}}&&=-\frac{\beta
ABe^{-Ar^{2}}}{2\pi}+\frac{\beta
A^{2}Br^{2}e^{-Ar^{2}}}{\pi}+\frac{\beta Ae^{-Ar^{2}}}{\pi
r^{2}}+\frac {\beta A^{2}e^{-Ar^{2}}}{2\pi}+\frac{3\beta
e^{-Ar^{2}}}{4 \pi r^{4}}\\\nonumber&+&\frac {\beta
Ae^{-Ar^{2}}}{4\pi r^{2}}-\frac{3\beta}{4\pi r^{4}}.
\end{eqnarray}
\begin{figure}
\center\epsfig{file=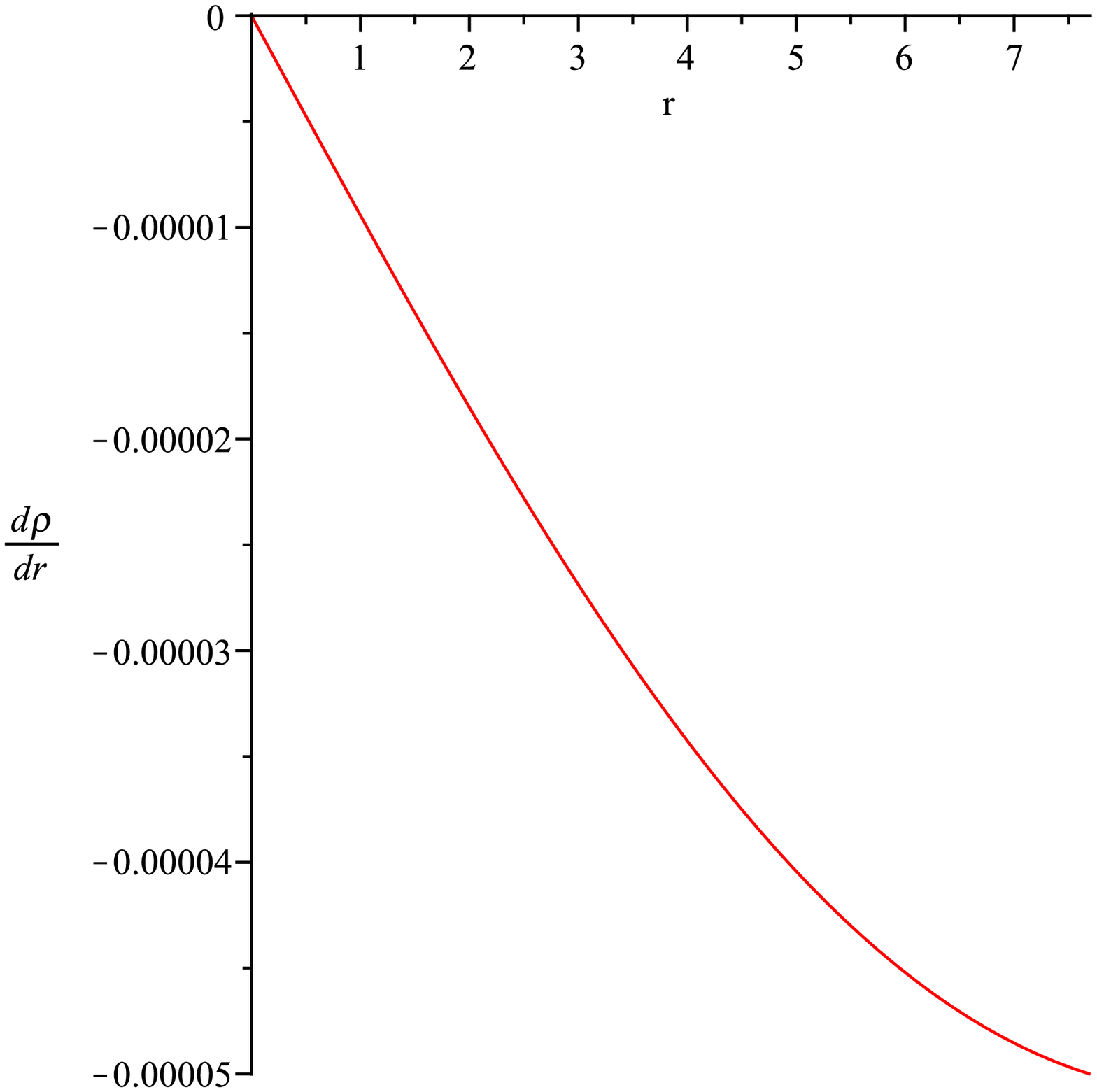, width=0.45\linewidth}
\epsfig{file=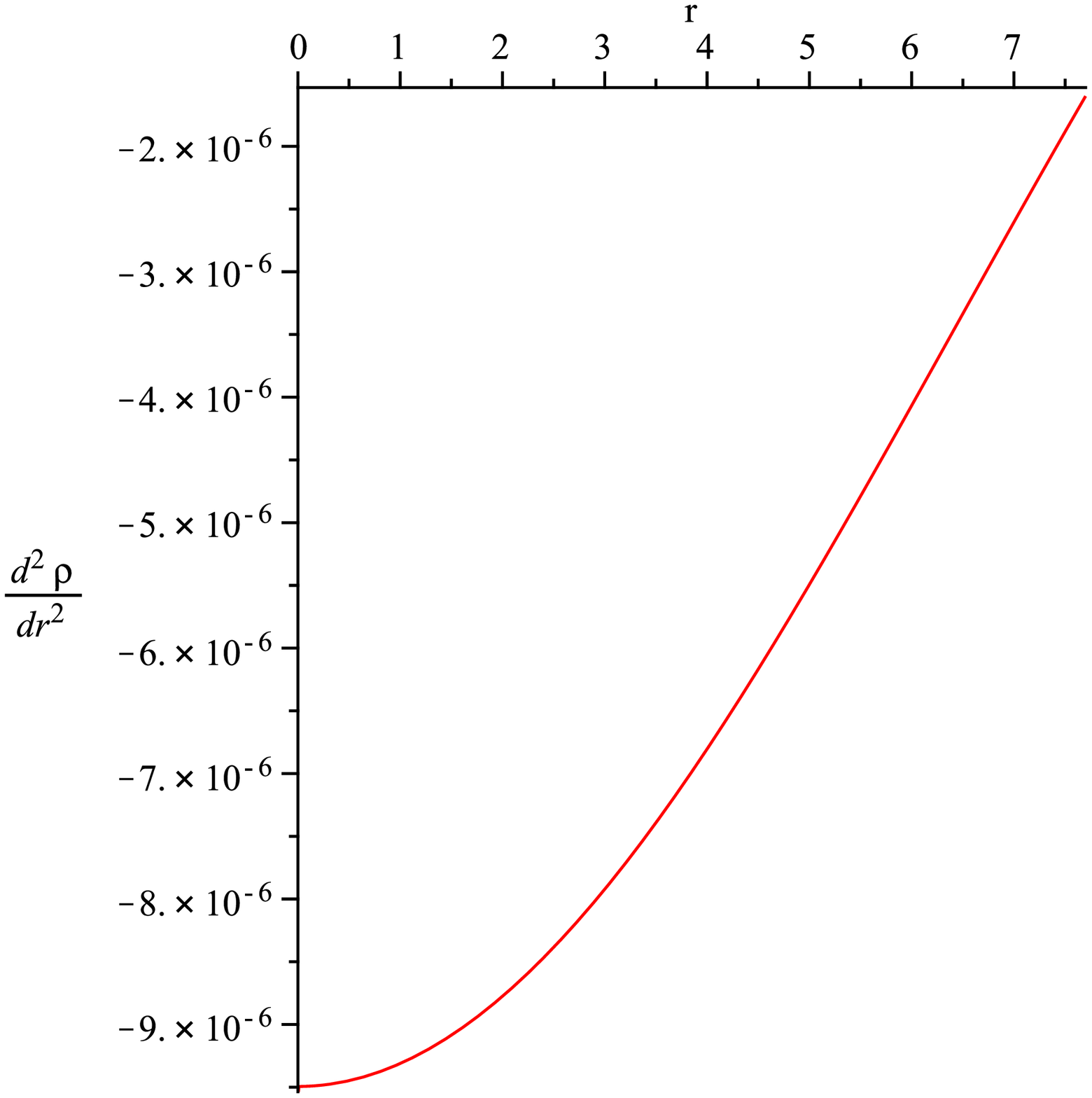, width=0.45\linewidth}
\epsfig{file=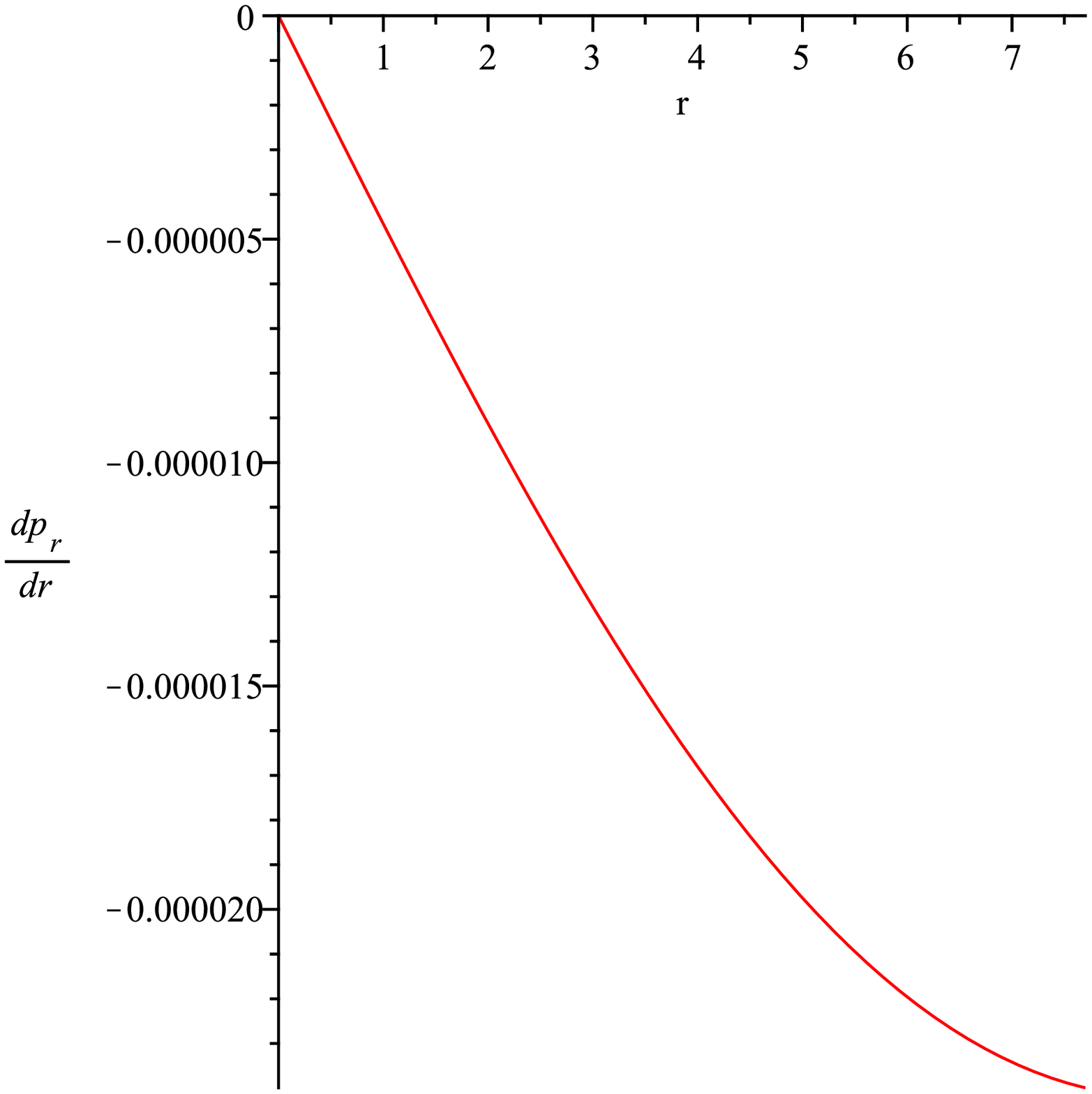, width=0.45\linewidth}\epsfig{file=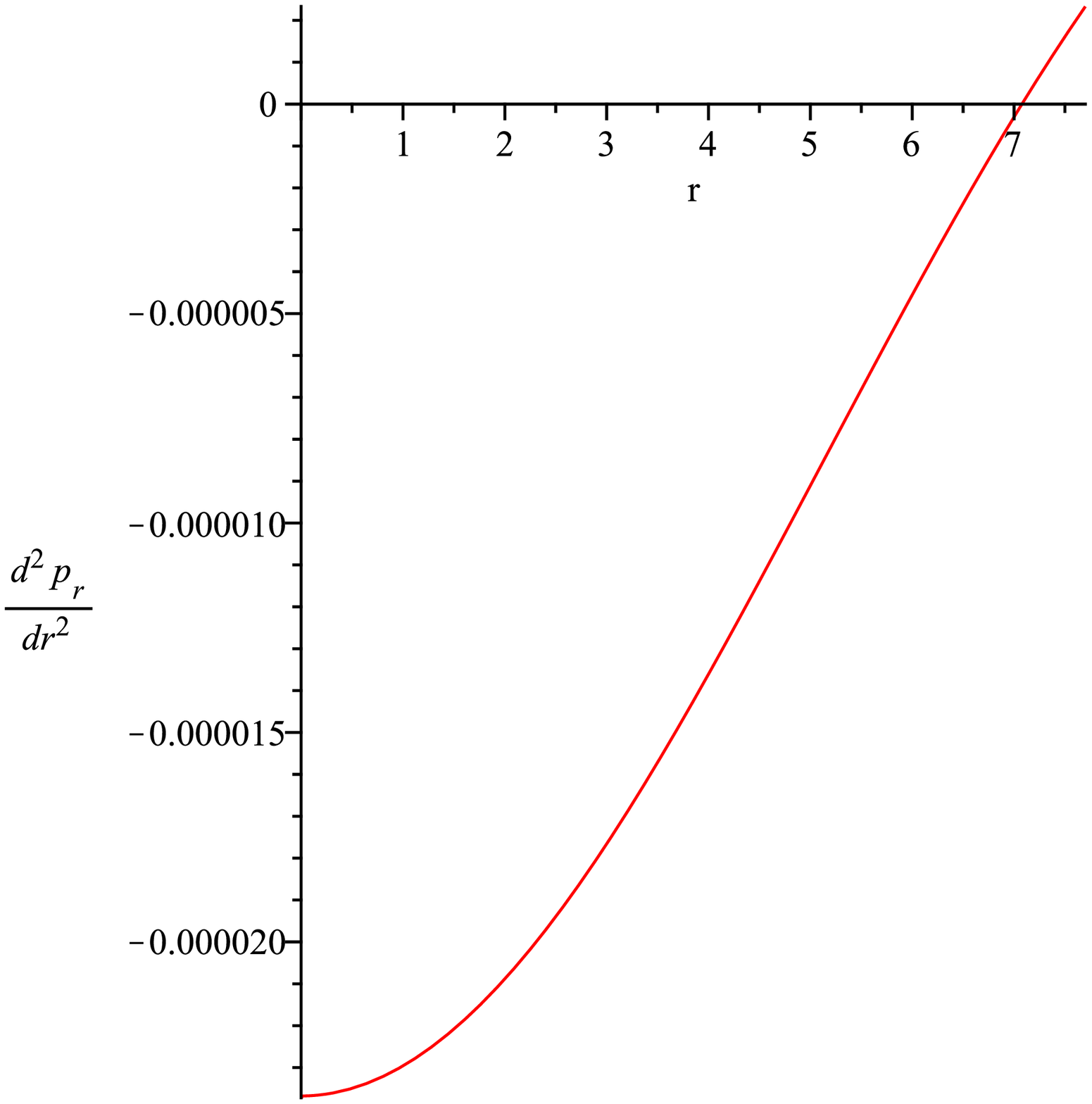,
width=0.45\linewidth}\caption{All these graphs has been plotted only
for the data of 4U 1820-30.}
\end{figure}
From figure \textbf{6}, we conclude that at $r=0$,
$\frac{d\rho}{dr}=0,~~ \frac{dp_r}{dr}=0$ and
$\frac{d^2\rho}{dr^2}<0,~~ \frac{d^2p_r}{dr^2}<0$, while at this
point $\rho$ and $p_r$ have maximum values (see figures \textbf{1}
and \textbf{2}). This implies that density and pressure have maximum
value at the center of the star ($r=0$). From figures \textbf{4} and
\textbf{5}, it can be see that effective EOS is given by
$0<\omega_i(r)<1$, ($i=r,t$) similar to normal matter distribution.
This indicates the fact that compact stars are composed of ordinary
matter and contribution of $f(T)$ terms. The measure of anisotropy,
$\Delta=\frac{2}{r}(p_{t}-p_{r})$ in this model is obtained as
follows:
\begin{equation}\label{4.5}
\Delta=\frac{\beta B^{2}re^{-Ar^{2}}}{4\pi}-\frac{\beta
ABre^{-Ar^{2}}}{4\pi}-\frac{\beta Ae^{-Ar^{2}}}{4\pi r}-\frac {\beta
e^{-Ar^{2}}}{4\pi r^{3}}+\frac{\beta}{2\pi r^{3}}.
\end{equation}
It is well known that anisotropy will be directed outward when
$p_{t}> p_{r}$ i.e., $ \Delta> 0$, and inward when $p_{t} < p_{r}$
i.e., $\Delta < 0$. Figure \textbf{7} shows that in this model a
repulsive (anisotropic) force would exists as $(\Delta> 0)$ (for
smaller values of $r$) which permits the formation of super massive
star, while for larger values of $r$, $\Delta=0$, where a star comes
to the equilibrium position.

\begin{figure}
\center\epsfig{file=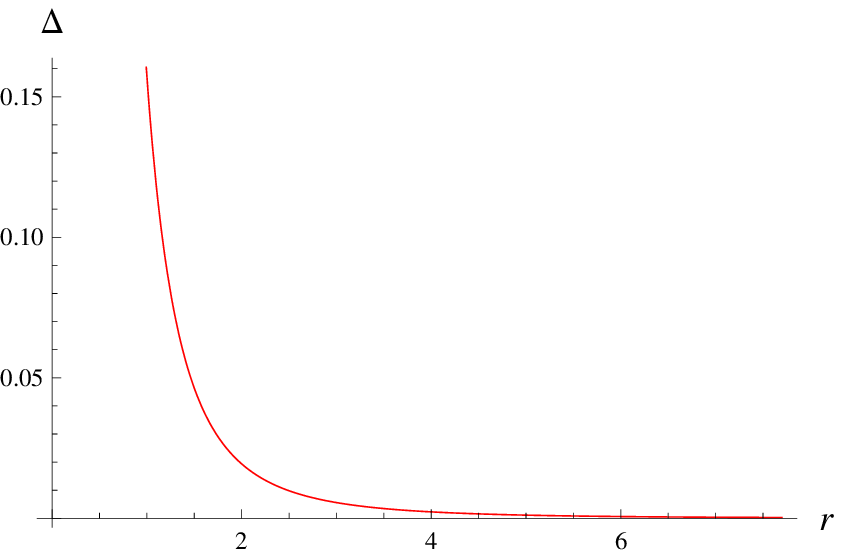, width=0.3\linewidth}
\epsfig{file=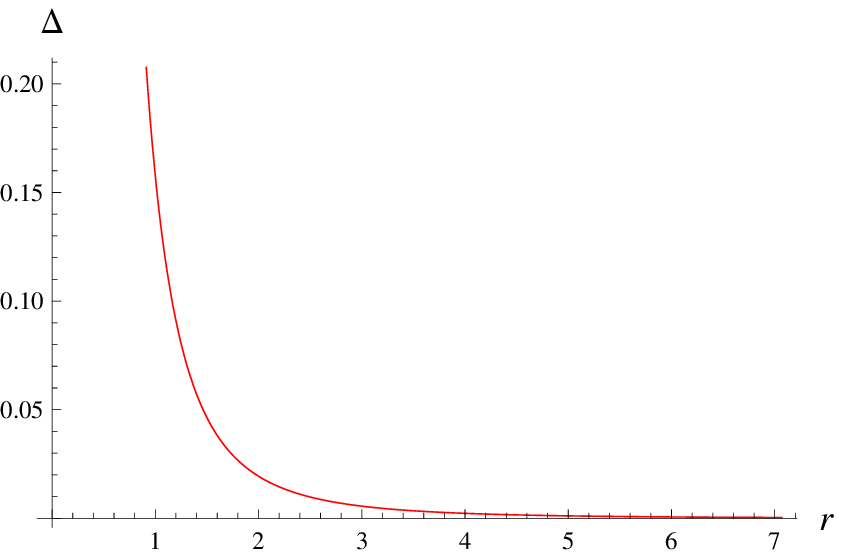, width=0.3\linewidth} \epsfig{file=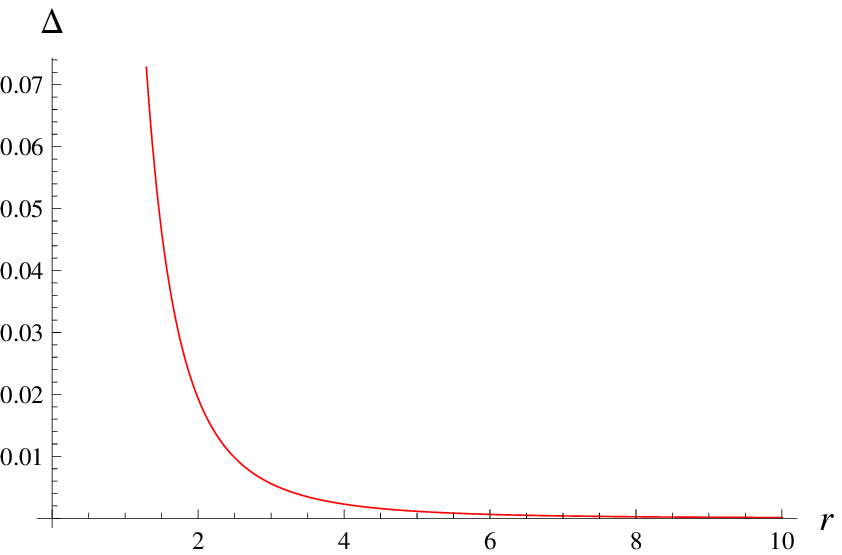,
width=0.3\linewidth}\caption{First, second and third graphs
represent the variation of anisotropy $\Delta$ for Strange star
candidate Her X-1, SAX J 1808.4-3658(SS1) and 4U 1820-30,
respectively.}
\end{figure}

\subsection{Matching Conditions}

In this section, we discuss the smooth matching of spacetime
(\ref{3.1}) to the vacuum exterior spherically symmetric metric
given by
\begin{equation}\label{4.6}
 ds^2=\left(1-\frac{2M}{r}\right)dt^2-\left(1-
 \frac{2M}{r}\right)^{-1}dr^2-r^{2}(d\theta^{2}+sin^{2}(\theta)d\phi^{2}).
\end{equation}
  The continuity of metric components $g_{tt}$,
  $g_{rr}$ and $\frac{\partial g_{tt}}{\partial r}$ at the boundary
  surface $r=R$ yield,
 \begin{eqnarray}\label{4.7}
  A&=&-\frac{1}{R^2}ln\left(1-\frac{2M}{r}\right),\\\label{4.8}
 B&=&\frac{M}{R^3}{{\left(1-\frac{2M}{r}\right)}^{-1}}\\
 C&=&ln\left(1-\frac{2M}{r}\right)-\frac{M}{R}{{\left(1-\frac{2M}{r}\right)}^{-1}}
\end{eqnarray}
For the values of $M$ and $R$ for a given star, the constants $A$
and $B$ can be specified as in table \textbf{1}.

\begin{table}[ht]
\caption{Values of constants for given Masses and Radii of Stars}
\begin{center}
\begin{tabular}{|c|c|c|c|c|c|}
\hline {Strange Quark Star}&  \textbf{ $M$} & \textbf{$R(km)$} &
\textbf{ $\frac{M}{R}$} &\textbf{ $A(km ^{-2})$}& \textbf{$B(km
^{-2})$}
\\\hline  Her X-1& 0.88$M_\odot$& 7.7&0.168&0.00749431669  &
$0.017062831$
\\\hline SAX J 1808.4-3658& 1.435$M_\odot$& 7.07&0.299& 0.010949753 &
$0.020501511$
\\\hline 4U 1820-30&2.25$M_\odot$& 10.0 &0.332&0.005715628647 &
$0.0101366226$
\\\hline
\end{tabular}
\end{center}
\end{table}

\subsection{Stability}

For this anisotropic model, the sound speeds are defined as
\begin{eqnarray}
\upsilon^{2}_{sr}&=&\frac{dp_{r}}{d\rho
}\equiv\frac{-2ABr^{4}e^{-Ar^{2}}-Ar^{2}e^{-Ar^{2}}-e^{-Ar^{2}}+1}{-2A^{2}r^{4}e^{-Ar^{2}}
+Ar^{2}e^{-Ar^{2}}+e^{-Ar^{2}}-1},\\
\upsilon^{2}_{st}&=&\frac{dp_{t}}{d\rho
}\equiv\frac{B^{2}r^{4}e^{-Ar^{2}}-AB^{2}r^{6}e^{-Ar^{2}}+4A^{2}Br^{6}
+A^{2}r^{4}e^{-Ar^{2}}}{Ar^{2}e^{-Ar^{2}}+e^{-Ar^{2}}-2A^{2}r^{4}e^{-Ar^{2}}-1}.
\end{eqnarray}
The above equations lead to
\begin{eqnarray}
\upsilon^{2}_{st}-\upsilon^{2}_{sr}=\frac{-2ABr^{4}
+B^{2}r^{4}-AB^{2}r^{6}+4A^{2}Br^{6}e^{Ar^{2}}+A^{2}r^{4}+(r^{2}
A+1)-e^{-Ar^{2}}}{Ar^{2}+1-2A^{2}r^{4}-e^{Ar^{2}}}\\\nonumber
\end{eqnarray}

Few years ago, Herrera (1992) published a new proposal to check the
stability of anisotropic gravitating source. Currently, this
technique is termed as cracking concept which states that if radial
speed of sound is greater than the transverse speed of sound in a
region then such a region is a potentially stable region, otherwise
unstable region. In our case, figure \textbf{8} indicates that there
is change of sign for the term ${{v^2}_{st}} -{{v^2}_{sr}}$ within
the particular configuration. Hence, we find that our strange star
model is unstable.

\begin{figure}
\center\epsfig{file=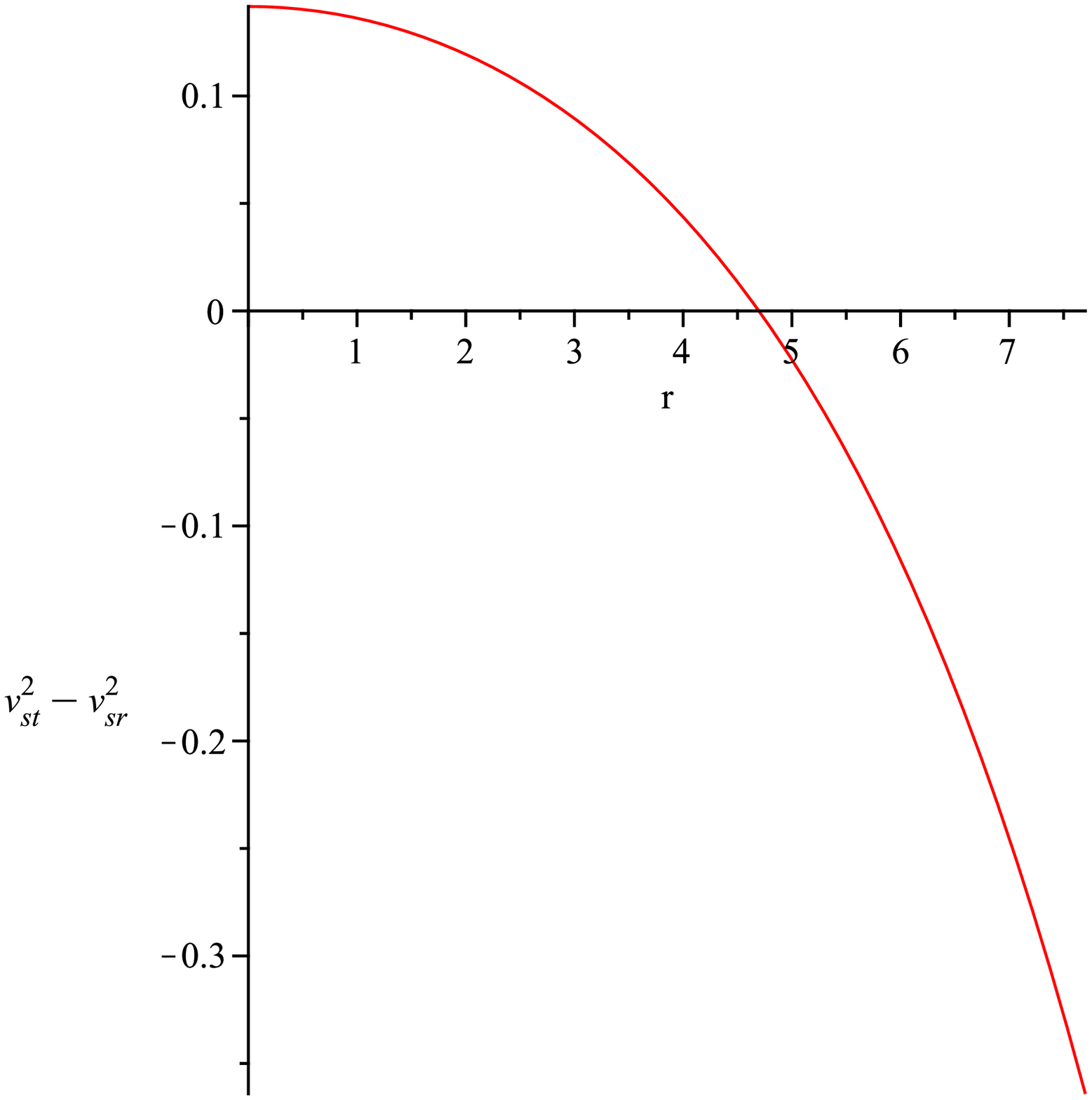, width=0.3\linewidth}
\epsfig{file=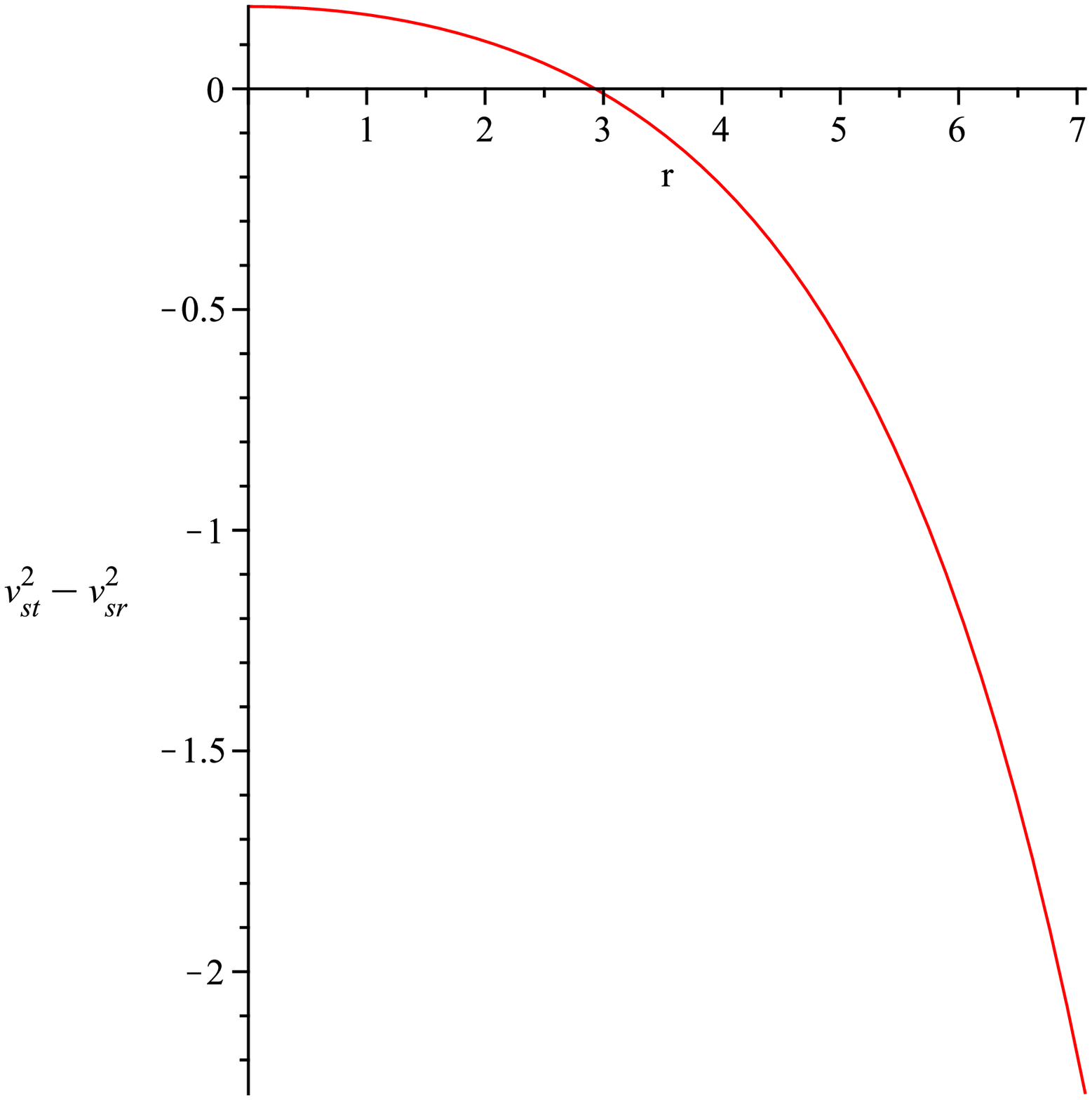, width=0.3\linewidth} \epsfig{file=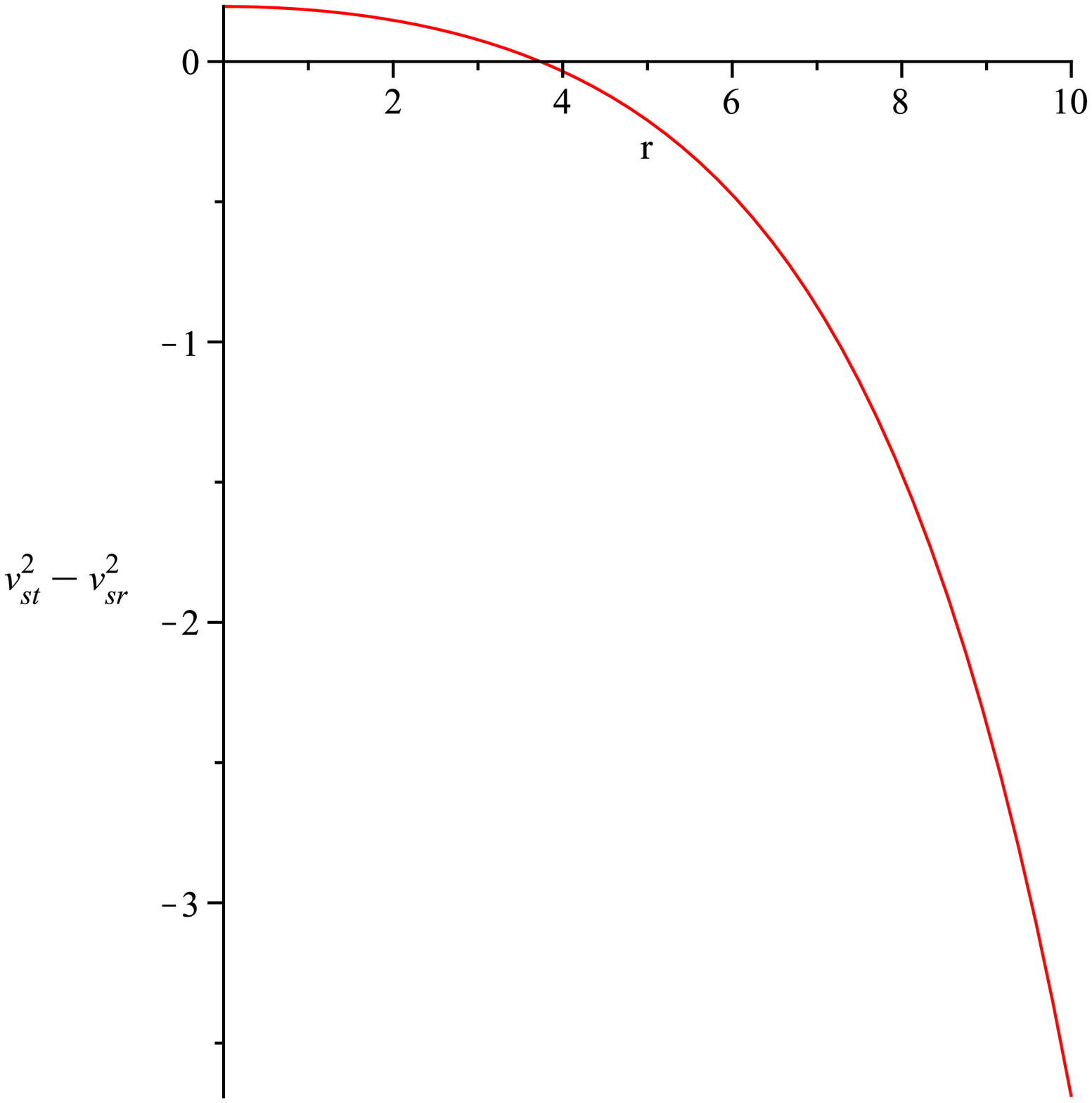,
width=0.3\linewidth}\caption{First, second and third graphs
represent the variation of ${{v^2}_{st}} -{{v^2}_{sr}}$ for Strange
star candidate Her X-1, SAX J 1808.4-3658(SS1) and 4U 1820-30,
respectively.}
\end{figure}

\begin{figure}
\center\epsfig{file=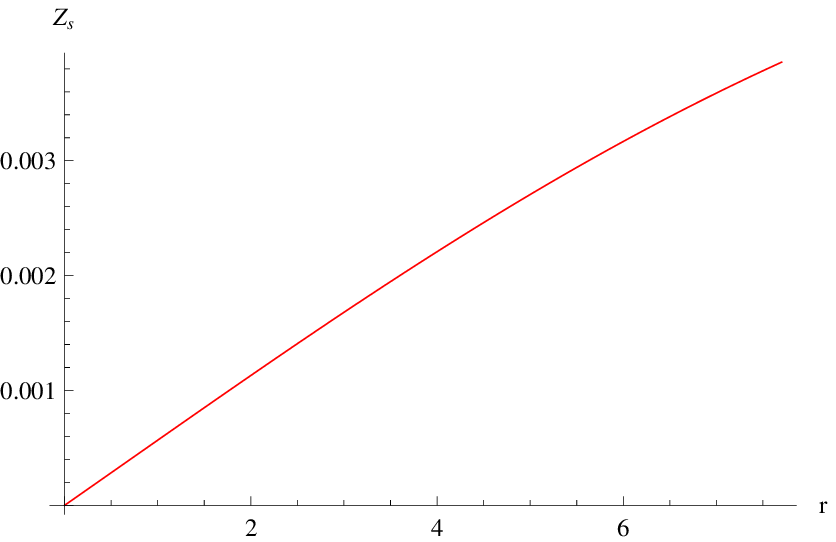, width=0.3\linewidth}
\epsfig{file=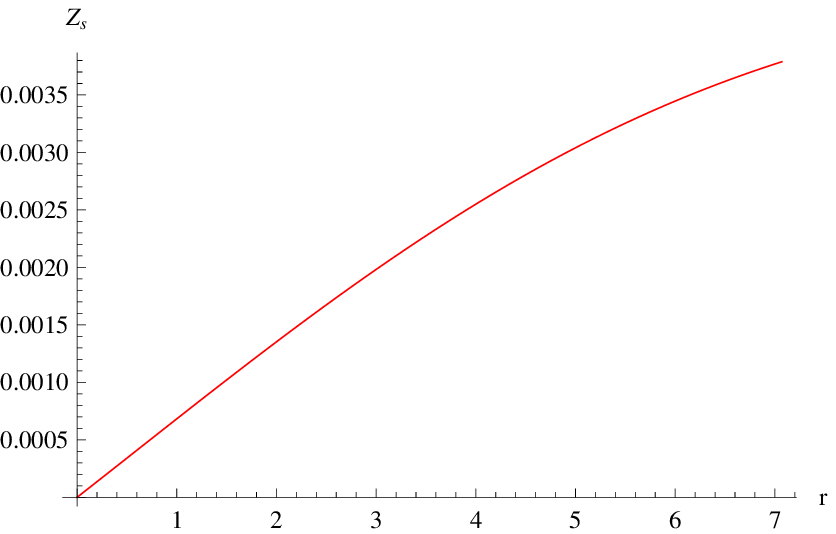, width=0.3\linewidth} \epsfig{file=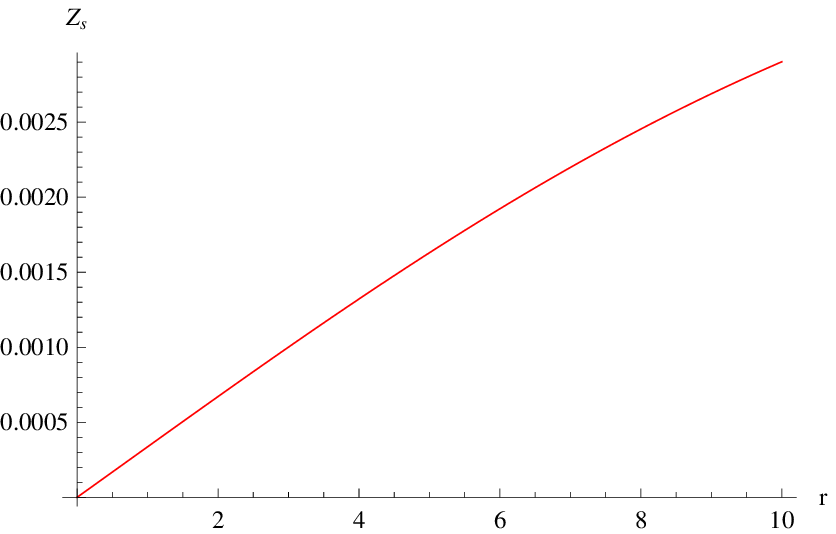,
width=0.3\linewidth}\caption{First, second and third graphs show the
evolution of surface redshift $Z_s$ for Strange star candidates Her
X-1, SAX J 1808.4-3658(SS1) and 4U 1820-30, respectively.}
\end{figure}

\subsection{Surface Redshift}

The compactness of the star is given by
\begin{equation}
u=\frac{M}{b}=\frac{2b\sqrt{A}\pi\beta+\sqrt{\pi}\beta(\pi-2A\pi
r^{2})erf[\sqrt{A}b]+b \pi r^{2}\beta_{1}\sqrt{A}}{4\pi\sqrt{A}}
\end{equation}

The surface redshift $(Z_{s})$ resulting from the compactness $u$ is
obtained as
\begin{equation}
1+Z_{s}=[1-2u]^\frac{-1}{2},
\end{equation}
where
\begin{equation}
1+Z_{s}=\left[1-\left(\frac{2b\sqrt{A}\pi\beta+\sqrt{\pi}\beta(\pi-2A\pi
r^{2})erf(\sqrt{A}b)+b \pi
r^{2}\beta_{1}\sqrt{A}}{4\pi\sqrt{A}}\right)\right]^{\frac{-1}{2}}.
\end{equation}
The maximum value of the surface redshift for the compact stars is
shown in figure \textbf{9}.

\section{Concluding Remarks}

Recently, there has been growing interest to study gravitational
field as the effect of torsion of the underlying geometry, which was
originally introduced in parallel to curvature description of
gravity. This concept was developed in several years as teleparallel
equivalence of GR. The black hole (BH) are extremely compact
astrophysical objects which store information about the entropy on
the BH horizon. In the modified TEGR, $f(T)$ gravity posses many
interesting feature. Recent observations from solar system orbital
motions in order to constrain $f(T)$ gravity have been made and
interesting results have been found.  The conditions for the
existence and non-existence of relativistic stars have been studied
in $f(T)$ gravity (Nojiri \& Odintsov 2007; Ferraro \& Fiorini
2007). It is important to model the compact stars in $f(T)$ using
the diagonal tetrad field.

In this paper, we have constructed the model of compact stars in
$f(T)$ gravity. The interior of the stars has been taken as static
spherically symmetric with anisotropic gravitating source. The
equations of motions have been obtained by using the diagonal tetrad
field. In this case the $f(T)$ appears as a linear function of $T$.
The explicit form of matter density, radial pressure, transverse
pressure and EOS parameters have been calculated. The anisotropy,
regularity and energy conditions have been discussed in detail. The
observed values of masses and radii of compact stars have been used
to specify the values of unknown constants of the interior metric.
By using the fist and second derivatives of density and pressures,
we have found that these quantities have maximum values at the
center and vanish at boundary.

It is clear from the figure \textbf{7} that in this model a
repulsive (anisotropic) force would exists as $(\Delta> 0)$ which
indicates the formation of more massive distributions. Using the
cracking concept, we have found that radial speed of sound is
greater than the transverse speed of sound in a region. It is clear
from figure \textbf{8} that there is change of sign for the term
${{v^2}_{st}} -{{v^2}_{sr}}$ within the specific configuration.
Hence, we conclude that our strange star model is unstable. The
range of surface redshift $Z_s$ for Strange star candidate Her X-1,
SAX J 1808.4-3658(SS1) and 4U 1820-30 is shown in figure \textbf{9}.

\end{document}